\documentclass[11pt]{article}%
\usepackage{amsmath,amsfonts,amssymb}%
\usepackage{color}
\usepackage{times,fourier}
\usepackage{fullpage}
\usepackage[colorlinks=true,citecolor=blue]{hyperref}
\usepackage{graphicx}
\graphicspath{{figs/}}
\definecolor{darkblue}{rgb}{0,0,0.7}

\providecommand{\U}[1]{\protect\rule{.1in}{.1in}}

\def\Complex{\mathbb{C}}
\def\Real{\mathbb{R}}
\advance\textwidth-4mm
\advance\hoffset2mm
\advance\textheight 6mm
\advance\voffset -2mm
\advance\parskip 2.5pt
\def\Re{\mathop{\rm Re}\nolimits}
\def\Im{\mathop{\rm Im}\nolimits}
\def\Wr{\mathop{\rm Wr}\nolimits}
\def\tr{\mathop{\rm Tr}\nolimits}
\def\diag{\mathop{\rm diag}\nolimits}
\catcode`\@ 11
\let\tru@sum=\sum
\def\sum{\mathop{\textstyle\tru@sum}\limits}
\def\@#1{{\mathbf{#1}}}

\catcode`\@ 12

\let\^=\hat
\let\~=\tilde
\let\_=\mathsf

\def\half{{\textstyle\frac12}}
\def\Xint#1{\mathchoice
{\XXint\displaystyle\textstyle{#1}}   {\XXint\textstyle\scriptstyle{#1}}   {\XXint\scriptstyle\scriptscriptstyle{#1}}   {\XXint\scriptscriptstyle\scriptscriptstyle{#1}}   \!\int}
\def\XXint#1#2#3{{\setbox0=\hbox{$#1{#2#3}{\int}$}
\vcenter{\hbox{$#2#3$}}\kern-.5\wd0}}
\def\ddashint{\Xint=}
\def\dashint{\Xint-}
\newcommand{\bs}{\begin{subequations}}
\newcommand{\es}{\end{subequations}}

\numberwithin{equation}{section}

\newcommand{\partialderiv}[3][]{\frac{\partial^{#1}#2}{\partial {#3}^{#1}}}
\allowdisplaybreaks

\def\Re{\mathop{\rm Re}\nolimits}
\def\Im{\mathop{\rm Im}\nolimits}
\def\tr{\mathop{\rm tr}\nolimits}
\def\arg{\mathop{\rm arg}\nolimits}

\def\Real{\mathbb{R}}

\def\U{\mathbb{U}}

\def\Complex{\mathbb{C}}
\let\<=\langle
\let\>=\rangle
\let\==\bar
\let\^=\hat
\let\~=\tilde
\let\.=\dot
\def\d{\mathrm{d}}
\def\e{\mathrm{e}}
\def\Wr{{\mathrm{Wr}}}
\def\tr{{\mathrm{tr}}}

\def\Res{\mathop{\rm Res}\nolimits}
\def\diag{\mathop{\rm diag}\nolimits}

\let\ge=\geqslant
\let\@=\mathbf

\let\eref=\eqref

\let\trueint=\int
\let\trueiint=\iint
\let\trueiiint=\iiint
\let\trueoint=\oint
\let\truesum=\sum
\let\trueprod=\prod
\def\@int#1{\mathchoice
{\@@int\displaystyle\textstyle{#1}}%
  {\@@int\textstyle\scriptstyle{#1}}%
  {\@@int\scriptstyle\scriptscriptstyle{#1}}%
  {\@@int\scriptscriptstyle\scriptscriptstyle{#1}}\!\int}
  \def\@@int#1#2#3{{\setbox0=\hbox{$#1{#2#3}{\textstyle\trueint}$}
  \kern0.4\wd0\vcenter{\hbox{$#2#3$}}\kern-0.666\wd0}}
\def\ddashint{\@int=}
\def\dashint{\@int-}
\def\int{\mathop{\textstyle\trueint}\limits}
\def\iint{\mathop{\textstyle\trueiint}\limits}
\def\iiint{\mathop{\textstyle\trueiiint}\limits}

\def\oint{\mathop{\textstyle\trueoint}\limits}
\def\oiint{\mathop{\circ\kern-1em\textstyle\trueint\kern-0.6em\trueint}\limits}
\def\sum{\mathop{\textstyle\truesum}\limits}
\def\prod{\mathop{\textstyle\trueprod}\limits}

\def\intinfty{\kern-0.2em\mathop{\textstyle\trueint}\limits_{\!-\infty\,}^{\,\,\infty\!}\kern-0.25em}
\def\txtintinfty{\kern-0.2em\mathop{\textstyle\trueint}\nolimits_{\!-\infty\,}^{\,\,\infty\!}\kern-0.25em}
\def\iintinfty{\mathop{\textstyle\trueiint}\limits_{\!\!-\infty\,\,}^{\,\,\infty\!}\kern-0.10em}
\def\iiiintinfty{\mathop{\textstyle\trueiiint}\limits_{\!\!-\infty\,\,}^{\,\,\infty\!}\kern-0.20em}
\makeatother

\def\Xint#1{\mathchoice
   {\XXint\displaystyle\textstyle{#1}}%
   {\XXint\textstyle\scriptstyle{#1}}%
   {\XXint\scriptstyle\scriptscriptstyle{#1}}%
   {\XXint\scriptscriptstyle\scriptscriptstyle{#1}}%
   \!\int}
\def\XXint#1#2#3{{\setbox0=\hbox{$#1{#2#3}{\int}$}
     \vcenter{\hbox{$#2#3$}}\kern-.5\wd0}}
\def\ddashint{\Xint=}
\def\dashint{\Xint-}

\def\[{\begin{equation}}
\def\]{\end{equation}}
\def\be{\begin{equation}}
\def\ee{\end{equation}}
\def\bse{\begin{subequations}}
\def\ese{\end{subequations}}
\def\bea{\begin{eqnarray}}
\def\eea{\end{eqnarray}}

\def\em{\endgroup}

\newdimen\figwdlt
\newdimen\figwdrt
\def\note[#1]{\marginpar{\color{red}[#1]}}

\begin{document}

\title{\bf On Maxwell-Bloch systems with inhomogeneous broadening and one-sided nonzero background}
\author{\large
Asela Abeya\,$^1$, Gino Biondini\,$^2$, Gregor Kova\v{c}i\v{c}\,$^3$ and Barbara Prinari\,$^{2,*}$
\\[1ex]
\normalsize\it
1: Department of Mathematics, SUNY Polytechnic Institute, Utica, NY, 13502\\
\normalsize\it
2: Department of Mathematics, State University of New York, Buffalo, NY, 14260\\
\normalsize\it
3: Department of Mathematical Sciences, Rensselaer Polytechnic Institute, Troy, NY, 12180\\
\normalsize\it
$^*$ Corresponding author. Email: bprinari@buffalo.edu
}
\date{\small\today}
\maketitle

\kern-4ex
\begin{abstract}
The inverse scattering transform is developed to solve the Maxwell-Bloch system of equations that describes
two-level systems with inhomogeneous broadening, in the case of optical pulses that do not vanish at infinity in the future.
The direct problem, which is formulated in terms of a suitably-defined uniformization variable,
combines features of the formalism with decaying as well as non-decaying fields.
The inverse problem is formulated in terms of a $2\times2$ matrix Riemann-Hilbert problem.
A novel aspect of the problem is that no reflectionless solutions can exist, and solitons are always accompanied by radiation. At the same time, it is also shown that, when the medium is initially in the ground state,
the radiative components of the solutions decay upon propagation into the medium,
giving rise to an asymptotically reflectionless states.
Like what happens when the optical pulse decays rapidly in the distant past and the distant future,
a medium that is initially excited decays to the stable ground state as $t\rightarrow \infty$ and for sufficiently large
propagation distances.
Finally, the asymptotic state of the medium and certain features of the optical pulse inside the medium
are considered, and the emergence of a transition region upon propagation in the medium is briefly discussed.
\end{abstract}

%%%%%%%%%%%%%%%%%%%%%%%%%%%%%%%%%%%%%%%%%%%%%%%%%%%%%%%%%%%%%%%%%%%%%%%%%%%%%%%%%%%%%%%%%%%%%%%%%%%%%%%%%%%%%%%%%%%%%
\section{Introduction}

Resonant interaction between light and optical media underlies several types of practical devices such as lasers and optical amplifiers~\cite{Boyd92,milonni10,shimoda86,siegman86}.
Typically, only narrow ranges of light colors interact resonantly with electron transitions between a small number of specific pairs of working energy levels in the active atoms~\cite{Allen87,Boyd92,ButcherCotter1990,HK1995,NewellMoloney1992}. Frequently, there is only one resonant transition and light is monochromatic, yet even this simple case produces a host of important physical effects such as: electromagnetically induced/self-induced transparency \cite{BEMS1990,PhysRevLett.66.2593,davis63,Jaynes63,harris:36,Slowlight99,PhysRevA.42.523,ISI:A1960ZQ06900019,mccall67,mccall69,PhysRevA.5.1634,PhysRevLett.29.1211}, superradiance and superflourescence \cite{BL1975,D1954,PhysRevLett.36.1035,HKSHG1979,PSV1979,PhysRevLett.30.309}, chaos and instabilities \cite{ISI:A1984TY33700007,ISI:A1985AWU8900005,PhysRevLett.58.2205,PhysRevLett.57.2804}, photon echo \cite{PhysRev.179.294,PhysRevLett.39.547,kurnit64,PS1968,ZM1982}, and remarkably even the slowing down of light to a tiny fraction of its speed in vacuum~\cite{Slowlight00,hau99,Milonni05,RVB2005-2,RVB2005-1}.

For many experimental and practical setups, a sufficient theoretical description of the interaction between light and an active optical medium is semi-classical, with the light described classically and the medium quantum-mechanically~\cite{Allen87}.   In the case of a finite number of resonant electron transitions, the quantum description reduces to a finite number of ordinary differential equations for the elements of the corresponding density matrix~\cite{ButcherCotter1990}.   When averaged over appropriate portions of the medium, this matrix renders a description of the macroscopic medium polarization as well as its average local level occupation~\cite{Boyd92,ButcherCotter1990}.   The large separation between the period(s) of the electromagnetic field oscillations corresponding to the color(s) of the light and the scale of its pulse width(s) further simplifies the theoretical description.   In particular, one can extract, and average over, the fast oscillations, and thus find the description only in terms of the slowly-varying envelopes corresponding to the evolution of the light intensity and phase~\cite{Allen87}.  Moreover, backscattering is neglected and thus only unidirectional propagation is assumed.  The resulting equations are called the Maxwell-Bloch equations (MBE), and are one of the fundamental models in modern nonlinear optics~\cite{Ablowitz74,L1969,Lamb71,Lamb74,risken:4662}.

The Maxwell-Bloch equations for two- and certain three-level media are completely integrable in the sense of possessing a Lax Pair (zero-curvature) representation~\cite{Ablowitz74}.
Integrability makes it possible to linearize exactly these equations via the
\textit{Inverse Scattering Transform} (IST)~\cite{ablowitz73b,BGK2003,CBA2014,GMZ1983,gabitov84,gabitov85,M1982,MN1986,steudel90},
and enables the use of various transformation methods to ``dress'' simple exact solutions into
more complicated and physically relevant ones~\cite{steudel88}.
At first,  only pulses whose intensity decays rapidly in both the distant past and the distant future were studied using IST.
Recently, however, the IST formalism was extended to optical pulses with \textit{nonzero background} (NZBG)
in both the distant past and the distant future, and pulses riding on top of continuous light beams
were also studied in~\cite{BGKL,Li_2018}.
In all these cases, the NZBG was assumed to be symmetric, i.e., approaching the same
nonzero amplitude as $t\rightarrow \pm \infty$.
In this work we investigate Maxwell-Bloch systems with \textit{one-sided NZBG}, corresponding to
light pulses riding on continuous waves that are in the process of either turning on or off.
Specifically, we consider one-sided boundary conditions with nonzero background in the distant future.

The integrable Maxwell-Bloch equations are special among integrable equations,
in that even the simplest problem involving them is an initial-boundary-value problem.
In some situations, the medium can be assumed to be semi-infinite, and ``prepared'' in the distant past (mathematically, in the limit as $t\rightarrow -\infty$)
in a (known) state characterized by assigned values for the distribution of atoms in the ground and excited states, and for the polarizations at every point.
In the case of a 2-level system, macroscopically the medium can be in: (i) a pure ground state (with all atoms in the lowest energy level); (ii) a pure excited state (i.e., a medium with a  complete ``population inversion'', with all the atoms in the excited state); (iii) a mixed state with an assigned fraction of atoms in each state (in this case, the medium exhibits nontrivial polarization fluctuations, encoded by the off-diagonal entries of the density matrix).
A light pulse is then injected into the medium at the origin, and it propagates to the right. The Maxwell-Bloch equations determine the optical pulse in each point of the medium at any given time (and, in particular, the residual optical pulse along the medium), as well as the final state of the entire medium after a long time (i.e., the asymptotics of the density matrix as $t\rightarrow +\infty$).
The state of the medium sample in the distant future cannot be assumed to be known, but it can be deduced from the state in the distant past (and the IST scattering data) by a rather sophisticated procedure first introduced in~\cite{Ablowitz74,gabitov85}.

As in all signaling-type problems, the role of the spatial and temporal variables in Maxwell-Bloch systems is reversed as compared to pure initial-value problems.
This is reflected in the IST treatment: the scattering and inverse scattering of the pulses takes place in time, and  the ``evolution'' is actually propagation along the medium~\cite{Ablowitz74,Lamb74}.
If all atoms in the medium are initially in the ground state, the propagation damps the ``continuous-radiation'' components of the solutions, and is thus not time-reversible~\cite{Ablowitz74}.
In this case, the response of the medium to an incident electric field, to which the medium is totally transparent and which
undergoes lossless propagation, is known as self-induced transparency.
The properties of the system change drastically when atoms are initially all in the excited state,
in which case the dynamics can give rise to the phenomenon called superfluorescence \cite{gabitov84,PhysRevLett.39.547,steudel90}.

Two classes of solutions of the MBE are naturally distinguished. If the medium is initially prepared in a pure state, and hence
does not exhibit polarization fluctuations as $t\rightarrow -\infty$, then the solution is completely determined by the incident pulse (in this work, $q(t,0)$). According to \cite{gabitov85}, such solutions are called ``causal'' solutions, in consideration of the fact that
if $q(t,0)=0$ for all $t<t_o$ for some $t_o\in \Real$, then one can show that $q(t,z)=0$ for all $t<t_o+z$, which means that the causal solution for a potential of finite range has a front which propagates into the medium with the velocity of light, in agreement with the notion of causality.
On the other hand, for a medium that is not initially in a pure state, the MBE admit nontrivial solutions even if $q(t,0)\equiv 0$, i.e., in the absence of an incident pulse. In this case, the solution is entirely determined by the polarization fluctuations as $t\rightarrow -\infty$, and, following the terminology introduced in
\cite{gabitov85,Zakharov80}, such solutions are called ``spontaneous'' solutions.  The solution of the MBE is in general a superposition of a causal and a spontaneous solution.

To avoid confusion, it is important to point out that the definition of ``causal'' solutions introduced in \cite{gabitov85}, which we also adopt in this work, is slightly different from the one recently introduced in \cite{LiMiller},
where the term ``causal'' was used to denote solutions for which the incident pulse $q(t,0)$ is identically zero for all $t<0$,
and the optical field $q(t,z)$  vanishes for all $t<0$ and $z\ge 0$.
(The reason for the latter definition is that the MBE are typically written in a comoving reference frame translating at the speed of light in vacuum -- including in this work, as well as in \cite{gabitov85,LiMiller} --
and therefore solutions with $t<0$ lie outside the light-cone frame and therefore are deemed to be unphysical.)
While the two definitions of ``causality'' agree when $q(t,0)=0$,
the first definition does not necessarily require the incident pulse to be of finite range.
A related issue is the question of whether solitons are to be considered unphysical in the Maxwell-Bloch system
because of the exponentially decaying tail, which extends to infinity in time.
This is true, strictly speaking.
However, it is still useful to include solitons in the description of two-level systems.
Indeed, the situation is exactly the same as for the nonlinear Schr\"odinger (NLS) equation.
The NLS equation is also written in a comoving frame like the Maxwell-Bloch system.
Still, solitons have been enormously fruitful objects in order to understand the properties of
optical systems governed by the NLS equation.
In fact, a truncated soliton (i.e., a soliton with its exponential tail ``chopped off'')
is described spectrally by a discrete eigenvalue plus a small radiative component.
Such an initial system configuration is certainly physical.
Moreover, since in the Maxwell-Bloch system in a stable system configuration
(e.g., when atoms are initially in the ground state) the radiation decays upon propagation,
and approximating the solution with solitons becomes increasingly more accurate upon propagation.

Another important feature of our work is that we do not limit ourselves to the sharp-line limit,
but instead deal explicitly with the presence of inhomogeneous broadening.
(Recall that, in optical media, the density matrix depends on the detuning from the exact quantum transition frequency due to the Doppler shift caused by
the thermal motion of the atoms in the medium,
and the associated MBE must account for the inhomogeneous broadening effect
by averaging over the range of detuning with the atomic velocity distribution function.
The sharp-line limit, sometimes also referred to as ``infinitely narrow line'',
corresponds to the limiting case in which the broadening function is taken to be a Dirac delta.)
As a byproduct, our treatment is not limited to media that are initially in a pure state.
Indeed, accounting for inhomogeneous broadening also allows considering a medium initially in a mixed state without requiring a compatible non-vanishing optical pulse
in the distant past, the latter becoming a constraint only in the sharp-line limit.
Besides the obvious physical relevance, including inhomogeneous broadening is also crucial to circumvent one of the
drawbacks highlighted in \cite{LiMiller},
namely the fact that no matter how fast the incident pulse $q(t,0)$ decays as $t\rightarrow +\infty$, after an infinitesimal propagation distance the optical pulse $q(t,z)$ always decays at a fixed slow rate as $t\rightarrow +\infty$.
This of course has consequences regarding the well-posedness of the IST, which in the case of zero boundary conditions requires that $q(\cdot,z)\in L^1(\Real)$ for all
$z\ge 0$, while even if this condition is imposed at $z=0$, it is generically violated for all $z>0$ (see Corollary 3 in \cite{LiMiller}).
Such slow decay of the optical field was also noted in \cite{gabitov85}, but in both cases it can be attributed to
the fact that neglecting inhomogeneous broadening
results in the second operator of the Lax pair exhibiting an essential singularity at the origin in the spectral plane.
With inhomogeneous broadening, there is no such essential singularity,
and the behavior of the reflection coefficient at the origin cannot play any role in inducing a slow decay of the optical field as $t\rightarrow +\infty$.

The outline of this work is as follows.
In section~\ref{s:prelims} we briefly present some preliminary mathematical facts about the problem,
to set up the framework for what follows.
In section~\ref{s:direct} we formulate the direct scattering problem of the IST,
including the Jost solutions, scattering matrix, symmetries, discrete eigenvalues, asymptotic behavior at the branch points, etc.
In section~\ref{s:inverse} we formulate the inverse problem both in terms of a Riemann-Hilbert problem from the left and one from the right, and we derive the trace formulae, which are needed to evaluate the propagation of the norming constants.
In section~\ref{s:asympdensity} we discuss the asymptotic values of the density matrix in the distant past and distant future,
and in section~\ref{s:propagation} we discuss the evolution (i.e., propagation) of the scattering data.
In section~\ref{s:lta} we use the IST formalism to briefly discuss the asymptotic behavior of the medium and of the optical pulse.
Finally, in section~\ref{s:conclusions} we offer some concluding remarks.

%%%%%%%%%%%%%%%%%%%%%%%%%%%%%%%%%%%%%%%%%%%%%%%%%%%%%%%%%%%%%%%%%%%%%%%%%%%%%%%%%%%%%%%%%%%%%%%%%%%%%%%%%%%%%%%%%%%%%
\section{Maxwell-Bloch equations, Lax pair and problem formulation}
\label{s:prelims}

Up to rescalings of dependent and independent variables, the MBE
that describe the propagation of an electromagnetic pulse $q(t,z)$ in a two-level medium characterized by a (real)
population density function $D(t,z,k)$ and (complex) polarization
fluctuation $P(t,z,k)$  for the atoms can be written
in dimensionless form \cite{Lamb71} as
\vspace*{-1ex}
\bse
\label{e:MBEcomp}
\begin{gather}
q_z(t,z)= - \int_{-\infty}^\infty P(t,z,k)\,g(k)\,dk\,, \\
P_t(t,z,k) -2ik P(t,z,k)=-2D(t,z,k) q(t,z)\,,
\\
D_t(t,z,k)=2 \Re \big[ q^*(t,z) P(t,z,k) \big]\,,
\end{gather}
\ese
where $z=z_{\mathrm{lab}}$ is the propagation distance, $t=t_{\mathrm{lab}}-z_{\mathrm{lab}}/c$ is a retarded time ($c$ being the speed of light in vacuum),
subscripts $z$ and $t$ denote partial differentiation,
the parameter $k$ is the deviation of the transition frequency of the atoms from its mean value,
$g(k)$ is the
so-called inhomogeneous broadening function, which describes the shape of the spectral line (the case $g(k)=\delta(k-k_o)$ corresponding to the so-called sharp-line limit, or infinitely narrow line),
and the asterisk $*$ denotes complex conjugation.

Introducing the matrix describing the optical field and the so-called density matrix of the medium,
\begin{equation}
Q(t,z)=\begin{pmatrix} 0 & q \\ -q^* & 0 \end{pmatrix}\,, \qquad \rho(t,z,k)=\begin{pmatrix} D & P \\ P^* & -D\end{pmatrix}\,,
\end{equation}
respectively, \eqref{e:MBEcomp} can be written compactly as
\vspace*{-1ex}
\bse
\label{e:MBE}
\begin{gather}
\rho_t=[ ik\sigma_3+Q,\rho]\,, \\
Q_z=-\frac{1}{2}\int_{-\infty}^{+\infty}[\sigma_3,\rho]\, g(k)\, dk
\end{gather}
\ese
where $[A,B]=AB-BA$ is the matrix commutator, and $\sigma_j$ for $j=1,2,3$ are the standard Pauli matrices, with $\sigma_3 = \diag(1,-1)$.
It was then shown in \cite{Lamb74,Ablowitz74} that \eref{e:MBE} are integrable, with a Lax pair given by
\bse
\label{e:Laxpair}
\begin{gather}
v_t= X v\,, \qquad
v_z=T v\,,
\label{e:Laxpair1}
\\
\noalign{\noindent with}
X(t,z,k) = ik\sigma_3 +Q,\qquad
T(t,z,k)=\frac{i\pi}{2}\mathcal{H}_k[\rho(t,z,\xi)g(\xi)]\,,
\label{e:Laxpair2}
\end{gather}
\ese
where $\mathcal{H}_k[f(\xi)]$ is the Hilbert transform, 
\vspace*{-1ex}
\begin{equation}
\mathcal{H}_k[f(\xi)]=\frac{1}{\pi}\dashint \frac{f(\xi)}{\xi-k}\d\xi\,,
\end{equation}
and the symbol $\dashint$ denotes the principal value integral over all reals.

In the following, the medium is assumed to be semi-infinite, i.e., $z\ge 0$, and ``prepared'' in the distant past
(i.e., as $t\rightarrow -\infty$)
in a (known) state characterized by assigned values for the distribution of atoms in the ground and excited states via $D(t,z,k)$, and for the polarization $P(t,z,k)$ at every point. The density matrix $\rho$ can be assumed without loss of generality to be traceless, and with determinant equal to $-1$ for all $z\ge 0$, so $D^2(t,z,k)+|P(t,z,k)|^2=1$.
An electromagnetic pulse $q(t,0)$ is then injected into the medium at the origin and it propagates into it ($z>0$).

The IST to solve the initial-value problem for the
above MBE with localized fields
[i.e., with $q(t,z)\to0$ as $t\to\pm\infty$]
was developed in \cite{Ablowitz74} in the case of an initially stable medium ($\lim_{t\rightarrow -\infty}D(t,z,k)=-1$) and in \cite{gabitov85} in the case of an arbitrary initial state of the medium.
The IST with a symmetric nonzero background (NZBG)
[i.e., $q(t,z)\to q_\pm (z)$ with $|q_+(z)|=|q_-(z)|=q_o$ as $t\to \pm \infty$]
was carried out in \cite{BGKL}. Here, we develop the IST for one-sided NZBG, namely:
\vspace*{-1ex}
\begin{gather}
q(t,z)\to
\begin{cases} 0 &t\to -\infty\,, \\
q_+(z) &t\to +\infty\,,
\end{cases}
\end{gather}
with $|q_+(z)|=A> 0$ for all $z\ge 0$.
This type of boundary conditions describes optical fields that start from zero and then never return back to zero,
remaining as a continuous wave (CW).
We point out that the same methodology can be used to study the situation when $q_+(z)\equiv 0$ and $q(t,z) \to q_-(z)\ne 0$ as $t\to -\infty$,
corresponding to optical fields that start as a CW and then get extinguished.
For brevity, however, we omit the details.
Note that, as in all earlier works and like in the study of optical fibers \cite{Agrawal2001},
in the formulation of the IST the set up is essentially that of a signaling problem,
in which the propagation distance $z$ is the evolution variable and $t$ is treated as a transverse variable.

\begin{figure}[b!]
\bigskip
\centerline{\includegraphics[width=0.30\textwidth]{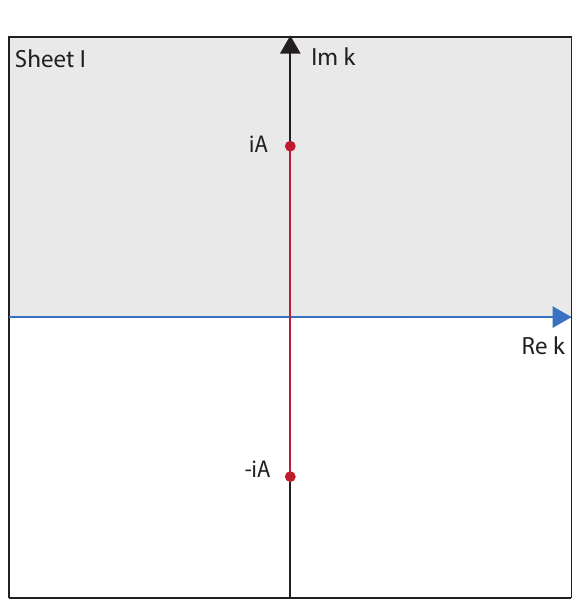}\qquad
\includegraphics[width=0.30\textwidth]{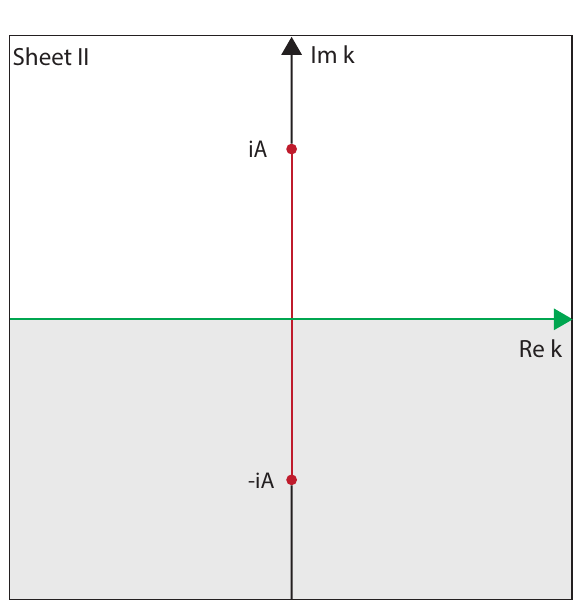}\qquad
\includegraphics[width=0.30\textwidth]{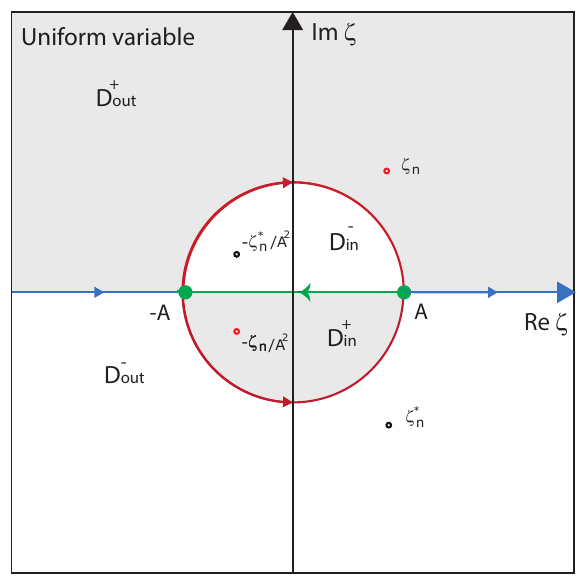}}
\bigskip
\caption{Left and center: the two-sheets of the
Riemann surface associated with $\lambda^2=k^2+A^2$. Right:
the complex plane for the uniformization variable
$\zeta=k+\lambda$. The grey regions ($D^+_\mathrm{in}$ and $D^+_\mathrm{out}$)
correspond to values of $\zeta$ for which $\Im \lambda>0$,
while the white regions ($D^-_\mathrm{in}$ and $D^-_\mathrm{out}$) correspond to
$\Im \lambda<0$.
The circle $\mathcal{C}$ (in red) corresponds to the cut $\Sigma$ on either sheet,
while $(-\infty,-A)\cup(A,+\infty)$
(blue) and $(-A,A)$ (green) correspond to the real $k$-axis on
sheets $\mathrm{I}$ and $\mathrm{II}$, respectively.}
\label{fig:zplaneonesided}
\end{figure}

The first half of the Lax pair~\eref{e:Laxpair1}, referred to as the scattering problem,
coincides with the scattering problem for the focusing NLS equation \cite{APT2004,ZS1972}.
As a result, the formulation of the direct problem is similar to that for the IST for the focusing NLS equation with one-sided nonzero background \cite{CM2015}.
On the other hand, as in the case of zero background \cite{Ablowitz74} and symmetric nonzero background \cite{BGKL},
the ``evolution'' of the scattering data for the MBE is substantially different and more complicated
than for the NLS equation, and also from the case of MBE with a symmetric NZBG.
Moreover, the formulation of the inverse problem in the present work is also substantially more involved than in the focusing NLS case \cite{CM2015}.

The asymptotic scattering problem as $t\to -\infty$ reduces to the one with zero background, while when $t\to +\infty$ it becomes
$v_t=X_+v$ where $X_+=ik\sigma_3+Q_+$, $Q_+(z)$ being $Q(t,z)$ with $q(t,z)$ replaced by its boundary value $q_+(z)$. The eigenvalues of $X_+$ are $\pm i \lambda$ with
$\lambda^2=k^2+A^2$,
where, as mentioned before, $A=|q_+|$. As
with the focusing NLS equation \cite{JMP55p031506}, to deal with the branching of the scattering parameter we consider the
two-sheeted Riemann surface defined by $\lambda(k)=(k^2+A^2)^{1/2}$ with a branch cut along the segment of the imaginary axis $[-iA,iA]$, and then introduce a uniformization variable defined by the conformal mapping \cite{FT1987}
\bse
\be
\zeta=k+\lambda(k),
\ee
which is inverted by the identities
\be
k = \half\left(\zeta-{A^2}/{\zeta}\right),
\qquad
\lambda = \zeta-k = \half\left(\zeta+{A^2}/{\zeta}\right).
\ee
\ese
The above transformation maps the
two-sheeted Riemann surface defined by $\lambda(k)$
onto a single complex $\zeta$-plane.
Specifically, one has the following results (cf.\ Fig.~\ref{fig:zplaneonesided}):
(i)
Sheet~I and sheet~II of the Riemann surface are mapped, respectively, onto the exterior
and the interior of the circle $\mathcal{C}$ of radius $A$.
(ii)
The branch cut $\Sigma$ on either sheet is mapped onto $\mathcal{C}$.
(iii)
The real $k$-axis on sheet~I and sheet~II $\mathrm{I}$
is mapped, respectively, onto $(-\infty,-A)\cup(A,+\infty)$ and $\left(-A,A\right)$.
(iv)
$\zeta(\pm iA)=\pm iA$ from either sheet,
while $\zeta(0^\pm_{\mathrm{I}})=\pm A$ and $\zeta(0^\pm_{\mathrm{II}})=\mp A$.
(v)
The upper half-plane $\Im k>0$ on sheet~I and sheet~II are mapped, respectively, into
the exterior and the interior of the circle $\mathcal{C}$ in the upper half-plane $\Im \zeta>0$,
whereas the lower half-plane $\Im k<0$ on sheet~I and sheet~II are mapped, respectively,
into the exterior and the interior of the circle $\mathcal{C}$ in the lower half-plane $\Im \zeta<0$.
It is therefore convenient to introduce the following regions in the complex $z$-plane:
\begin{equation*}
D^+=\{\zeta\in\mathbb{C}:(|\zeta|^2-A^2)\,\Im \zeta>0\},\quad
D^-=\{\zeta\in\mathbb{C}:(|\zeta|^2-A^2)\,\Im \zeta<0\},
\end{equation*}
which correspond to the regions where $\Im \lambda>0$ and $\Im \lambda<0$,
respectively, on either sheet.
The complex $\zeta$-plane is then partitioned into four regions: the upper/lower half $\zeta$-plane outside
the circle $\mathcal{C}$, denoted as $D^{\pm}_{\mathrm{out}}$, respectively,
and the lower/upper half $\zeta$-plane inside the circle $\mathcal{C}$ denoted
as $D^{\pm}_{\mathrm{in}}$, respectively. In the following, we will
also denote with $\mathcal{C}^\pm$ the upper and lower semicircles of
radius $A$, respectively.

Note that in general the density matrix $\rho(t,z,k)$ is only defined for $k\in \Real$. In terms of the uniformization variable $\zeta$, we can evaluate it for
all $\zeta \in \Real$, specifically for $|\zeta|>A$ on sheet I, and for $-A<\zeta<A$ on sheet II, where obviously $\rho(t,z,-A^2/\zeta)=\rho(t,z,
\zeta)$ since $\rho$ is a single-valued function of $k$. But we cannot assume $\rho$ is defined off the real $k$ axis.

Note also that, in analogy with what happens for the symmetric NZBG, the background solution should reduce to $\rho=h_3 \sigma_3$ with $h_3=\pm 1$ in the limit $A\to 0$,
and to maintain consistency one needs to choose the opposite sign of $h_3$ on sheet II.

%%%%%%%%%%%%%%%%%%%%%%%%%%%%%%%%%%%%%%%%%%%%%%%%%%%%%%%%%%%%%%%%%%%%%%%%%%%%%%%%%%%%%%%%%%%%%%%%%%%%%%%%%%%%%%%%%%%%%
\section{Direct scattering problem}
\label{s:direct}

\subsection{Jost solutions, analyticity and scattering matrix}

As usual, the direct problem in the IST consists in mapping the solution of the MBE into a suitable set of scattering data.
And, as usual, this is done by introducing Jost eigenfunctions,
which are solutions of the scattering problem with prescribed exponential asymptotic behavior at infinity.
Specifically, in light of the asymptotic behaviors of the  scattering problem as $t\to\infty$ discussed above,
we define the Jost eigenfunctions as
\bse
\label{e:Jostsols}
\begin{gather}
\Phi(t,z,\zeta) = \left(\bar{\phi}(t,z,\zeta)\,\,, \phi(t,z,\zeta) \right)
  = I_2\e^{ik(\zeta)t\sigma_3}\,(1 + o(1)), \qquad t\to -\infty\,, \\
\Psi(t,z,\zeta) = \left(\psi(t,z,\zeta)\,,\,\bar{\psi}(t,z,\zeta) \right)
  = Y_+(\zeta,z) \,\e^{i\lambda(\zeta)t\sigma_3}\, (1+o(1)), \qquad t\to +\infty\,,
\end{gather}
\ese
where $I_2$ is the $2\times2$ identity matrix,
\be
Y_+(\zeta,z) = I_2+\frac{i}{\zeta}\sigma_3Q_+(z)\,,
\ee
and $\=\phi$, $\phi$, $\psi$ and $\=\psi$ denote the first and second columns of $\Phi$ and $\Psi$, respectively.
(Note that in this work the overbar is not used to denote Schwarz conjugation.)
Then $\Phi(t,z,\zeta)$ is defined for $k(\zeta)\in \Real$
[i.e., for $\zeta \in \Real$],
while
$\Psi(t,z,\zeta)$ is defined for $\lambda(\zeta)\in \Real$
[i.e., for $\zeta\in\Real\cup \mathcal{C}$].
One can introduce modified eigenfunctions by removing the asymptotic exponential oscillations, and integrate the resulting ODEs for the modified eigenfunctions to obtain integral equations.
Standard Neumann series iterations on these Volterra integral equations allow one to prove that if the potential $q(t,z)$ satisfies the condition
\be
\qquad \int_{\Real}(1+|t|)|q(t,z)-q_+(z)H(t)|\,\d t<\infty
\label{e:H1}
\ee
where $H(t)$ is the Heaviside function, then $\psi(t,z,\zeta)$ is an analytic function of $\zeta$ in $D^+$ and continuous up to the boundary of $D^+$,
while $\bar{\psi}(t,z,\zeta)$ is analytic in $D^-$ and continuous up to $\partial D^-$.
Similarly, $\phi(t,z,\zeta)$ and $\bar{\phi}(t,z,\zeta)$ are analytic in $\Complex^+$ and $\Complex^-$, respectively,
and continuous for $z\in \Real$.

As usual, we define the continuous spectrum of the scattering problem as the set of values of $k$
where all four eigenfunctions are simultaneously defined.
Unlike what happens with symmetric NZBG
(for which the continuous spectrum includes $\mathcal{C}$),
here the continuous spectrum is limited to the real $\zeta$-axis.
Abel's theorem implies that, for
any matrix solution $v$ of the scattering problem,
$\partial_t(\det v)=0$.
In addition, for all $\zeta\in\Real$, $\lim_{t\to -\infty}\Phi(t,z,\zeta)\,\e^{-ik(\zeta)t\sigma_3}=I_2$
and $\lim_{t\to +\infty}\Psi(t,z,\zeta)\,\e^{-i\lambda(\zeta)t\sigma_3}=Y_+(\zeta,z)$.
Hence,
$\forall t,\zeta\in \Real$ we have
$\det \Phi(t,z,\zeta)=1$ and
$\det \Psi(t,z,\zeta)=\det Y_+(\zeta,z)$.
Thus, for all $\zeta\in \Real$, both
$\Phi$ and $\Psi$ are two fundamental matrix solutions of the scattering problem,
and one can express one set of eigenfunctions in terms of the other one:
\begin{subequations}
\label{e:scattmatrix}
\begin{align}
\label{scatt_mat_z}
\Psi(t,z,\zeta)=\Phi(t,z,\zeta)\, S(\zeta,z)\,, \qquad
S(\zeta,z)=\begin{pmatrix} a(\zeta,z)& \bar{b}(\zeta,z)\\
b(\zeta,z)& \bar{a}(\zeta,z)\end{pmatrix}\,,\quad \zeta\in \Real\,, \\
\label{scatt_mat_z2}
\Phi(t,z,\zeta)=\Psi(t,z,\zeta)\, S^{-1}(\zeta,z)\,, \qquad
S^{-1}(\zeta,z)=\begin{pmatrix}\bar{c}(\zeta,z)&d(\zeta,z)\\
\bar{d}(\zeta,z)&c(\zeta,z)\end{pmatrix}\,,\quad \zeta \in \Real\,,
\end{align}
\end{subequations}
where $S(\zeta,z)$ is the scattering matrix, whose entries are referred to as the scattering coefficients.
Note that unlike the case of symmetric NZBG, here the scattering matrix is not unimodular, since
\eref{e:scattmatrix} implies $\det S = \det\Psi$,  i.e., explicitly:
\begin{equation}
\det S(\zeta,z)={2\lambda(\zeta)}/{\zeta} = 1+A^2/\zeta^2 \,.
\label{e:detS}
\end{equation}
As usual, one can express the scattering coefficients as Wronskians of the Jost solutions:
\begin{subequations}
\label{Wrapz}
\begin{align}
\label{Wrapaz}
a(\zeta,z)=\Wr(\psi(t,z,\zeta),\phi(t,z,\zeta)), \qquad
\bar{a}(\zeta,z)=\Wr(\phi(t,z,\zeta),\bar{\psi}(t,z,\zeta)), \\
\label{Wrapbz}
b(\zeta,z)=\Wr(\bar{\phi}(t,z,\zeta),\psi(t,z,\zeta)), \qquad
\bar{b}(\zeta,z)=\Wr(\bar{\psi}(t,z,\zeta),\phi(t,z,\zeta)).
\end{align}
\end{subequations}
In turn, \eref{Wrapz} imply that:
(i) $a(\zeta,z)$ is analytic in $D^+_{\mathrm{out}}$ and continuous for $\zeta\in\Real \cup \mathcal{C}^+ \setminus\{iA \}$;
(ii) $\bar{a}(\zeta,z)$ is analytic in $D^-_{\mathrm{out}}$ and continuous for $\zeta\in \Real \cup \mathcal{C}^-\setminus\{-iA\}$;
(iii) $b(\zeta,z)$ is analytic in $D^+_{\mathrm{in}}$ and continuous for $\zeta\in \Real \cup \mathcal{C}^-\setminus\{-iA\}$;
(iv) $\bar{b}(\zeta,z)$ is analytic in $D^-_{\mathrm{in}}$ and continuous for $\zeta\in\Real \cup \mathcal{C}^+\setminus\{iA\}$.

The first set of reflection coefficients that will be needed in the inverse problem,
which we refer to as the reflection coefficients from the left, is given by:
\bse
\begin{equation}
\label{reflcoeffz}
r_-(\zeta,z)=\frac{b(\zeta,z)}{a(\zeta,z)}\,,\qquad
\bar{r}_-(\zeta,z)=\frac{\bar{b}(\zeta,z)}{\bar{a}(\zeta,z)}\,, \qquad \zeta\in \Real\,.
\end{equation}
Similarly, one can define reflection coefficients from the right in terms of
the entries of $S^{-1}(\zeta,z)$:
\begin{equation}
r_+(\zeta,z)=\frac{d(\zeta,z)}{c(\zeta,z)}\equiv -\frac{\bar{b}(\zeta,z)}{a(\zeta,z)}\,,\qquad
\bar{r}_+(\zeta,z)=\frac{\bar{d}(\zeta,z)}{\bar{c}(\zeta,z)}\equiv-\frac{b(\zeta,z)}{\bar{a}(\zeta,z)}\,, \qquad \zeta\in \Real\,.
\label{e:reflcoeff_+}
\end{equation}
\ese
Note that, unlike $r_-(\zeta,z)$ and $\bar{r}_-(\zeta,z)$, the reflection coefficients from the right are such that
$r_+(\zeta,z)$ is also defined on $\mathcal{C}^+$ (except, possibly, at $\zeta=iA$, where $\bar{b}(\zeta,z)$ and $a(\zeta,z)$ might have a pole), and $\bar{r}_+(\zeta,z)$ is also defined on $\mathcal{C}^-$ (except, possibly, at $\zeta=-iA$, where $b(\zeta,z)$ and $\bar{a}(\zeta,z)$ might have a simple pole, see \cite{JMP55p031506,CM2015}).
One can also write an integral representation of the scattering matrix $S(\zeta,z)$, which is analogous to the one with symmetric NZBG
with $Y_-\equiv I_2$.
Since such a representation will not be used, however, it is omitted for brevity.

\subsection{Symmetries of eigenfunctions and scattering coefficients}
\label{s:symmetries}

The scattering problem admits two nontrivial symmetries:
$(k,\lambda) \mapsto (k^*,\lambda^*)$ (i.e., switching between the upper and lower half $k$-planes)
and $(k,\lambda)\mapsto (k,-\lambda)$ (i.e., switching between opposite sheets).
In terms of the uniformization variable $\zeta$,
these correspond to the maps
$\zeta \mapsto \zeta^*$ (i.e., upper/lower half $\zeta$-plane) and
$\zeta \mapsto -A^2/\zeta$ (outside/inside the circle $\mathcal{C}$).

Regarding the first involution, the boundary conditions~\eref{e:Jostsols} yield the following symmetries for the Jost solutions
and, in turn, for the scattering coefficients and reflection coefficients:
\bse
\label{symmJost1zvar}
\begin{gather}
\bar{\psi}^*(t,z,\zeta^*)=-i\sigma_2\psi(t,z,\zeta),\qquad \zeta\in D^+\cup \mathcal{C}\cup\Real\,,\\
\psi^{*}(t,z,\zeta^*)=i\sigma_2\bar{\psi}(t,z,\zeta),\qquad \zeta\in D^-\cup \mathcal{C} \cup\Real \,,
\label{symmJost1azvar}
\\
\phi^{*}(t,z,\zeta^*)=-i\sigma_2 \bar{\phi}(t,z,\zeta),\qquad \zeta\in  \Complex^- \cup \Real \,,\\
\bar{\phi}^{*}(t,z,\zeta^*)=i\sigma_2\phi(t,z,\zeta),\qquad \zeta\in \Complex^+\cup \Real  \,.
\label{symmJost1bzvar}
\\
\bar{a}^*(\zeta^*,z)=a(\zeta,z),\qquad
\zeta\in D^+_{\mathrm{out}}\cup \mathcal{C}^+\cup \Real\,,
\label{symmSCAT1azvar}
\\
\bar{b}^*(\zeta^*,z)=-b(\zeta,z),\qquad
\zeta\in D^-_{\mathrm{in}}\cup \mathcal{C}^-\cup\Real\,.
\\
\bar{r}_-^*(\zeta^*,z)=-r_-(\zeta,z)\,, %\qquad
\quad \zeta\in\Real\,,
\label{e:1stsymmr_-}
\\
\bar{r}_+^*(\zeta^*,z)=-r_+(\zeta,z)\,,\quad
\zeta\in \mathcal{C}^+ \cup \Real\setminus\{iA\}\,.
\label{e:1stsymmr_+}
\end{gather}
\ese
Using the above symmetries, \eqref{e:detS} can then be written as
\begin{equation}
\label{e:a2-b2}
|a(\zeta,z)|^2+|b(\zeta,z)|^2=2\lambda /\zeta \qquad \zeta \in \Real\,, z\ge 0.
\end{equation}

For the second involution, since
$\lambda(-A^2/\zeta)=-\lambda(\zeta)$ and $k(-A^2/\zeta)=k(\zeta)$, taking into account the boundary
conditions \eref{e:Jostsols}, one can easily establish the following additional
symmetry relations:
\bse
\label{e:Symm2Jost}
\begin{gather}
\label{Symm2psizvar1}
\bar{\psi}(t,z,\zeta)=\frac{iq_+(z)}{\zeta}\psi(t,z,-A^2/\zeta),\qquad
\zeta\in D^-\cup \mathcal{C} \,,\\
\psi(t,z,\zeta)=\frac{iq_+^*(z)}{\zeta}\bar{\psi}(t,z,-A^2/\zeta) ,\qquad
\zeta\in D^+\cup \mathcal{C}\,,\label{Symm2psizvar2} \\
\phi(t,z,\zeta)=\phi(t,z,-A^2/\zeta) ,\quad \zeta\in \Real\cup \mathcal{C}^+\,, \\
\bar{\phi}(t,z,\zeta)=\bar{\phi}(t,z,-A^2/\zeta),\qquad
\zeta\in \Real\cup \mathcal{C}^-\,.
\\
\label{Symm2SCATTzvar}
a(\zeta,z)=\frac{iq_+^*(z)}{\zeta} \bar{b}(-A^2/\zeta,z),\qquad
  \zeta\in D^+_{\mathrm{out}}\cup\Real \cup \mathcal{C}^+\,, \\
\label{Symm2SCATTbbarzvar}
\bar{a}(\zeta,z)=\frac{iq_+(z)}{\zeta}b(-A^2/\zeta,z),\qquad
  \zeta\in D^-_{\mathrm{out}}\cup\Real\cup \mathcal{C}^-\,.
\\
\label{Symm2SCATTczvar} c(\zeta,z)=-\frac{i\zeta}{q_+(z)}d(-A^2/\zeta,z),\qquad
  \zeta\in D^+_{\mathrm{out}}\cup\Real\cup \mathcal{C}^+\,, \\
\label{Symm2SCATTcbarzvar}
\bar{c}(\zeta,z)=-\frac{i\zeta}{q_+^*(z)}\bar{d}(-A^2/\zeta,z) ,\qquad
  \zeta\in D^-_{\mathrm{out}}\cup\Real\cup \mathcal{C}^-\,.
\end{gather}
\ese
In matrix form, the above symmetry can be written as
\bse
\begin{gather}
\Psi(t,z,\zeta)=\frac{i}{\zeta}\Psi(t,z,-A^2/\zeta)\sigma_3 Q_+(z)\,, \qquad
\Phi(t,z,\zeta)=\Phi(t,z,-A^2/\zeta)\,.
\\
S(\zeta,z) = \frac{i}{\zeta} S(-A^2/\zeta,z) \sigma_3 Q_+(z)\,.
\end{gather}
\ese
Note that, unlike what happens for the focusing NLS equation, the above symmetries depend explicitly on $z$.
The explicit $z$-dependence will be determined in section~\ref{s:propagation}, where we discuss the propagation
of the background as well as that of all relevant quantities in the inverse scattering formalism.

Combining the first and second involutions above also yields the following symmetry for the scattering matrix:
\be
S^*(\zeta^*,z) = \frac i{\zeta} \sigma_2 S(-A^2/\zeta,z) \sigma_3 Q_+(z)\sigma_2\,.
\ee
As a direct consequence we have the following symmetry for the reflection coefficients from the left:
\be
\label{e:symmr_-}
r_-^*(\zeta^*, z) \,r_-(-A^2/\zeta,z) = \=r_-(\zeta, z) \,\=r_-^*(-A^2/\zeta^*,z) = -1\,,\quad \zeta \in \Real\,
\ee
On the other hand, the second symmetry implies that the reflection coefficients from the right satisfy the symmetry:
\be
r_+(\zeta, z) \,r_+(-A^2/\zeta,z) = \e^{2i\arg q_+(z)}
\,,\quad \zeta \in \Real\,.
\label{e:symforr_+}
\ee

Importantly, \eqref{e:symmr_-} and~\eqref{e:symforr_+} imply that $r_\pm(\zeta,z)\ne0$ for all $\zeta\in\Real$.
This is significant for two reasons:
(i)~It means that no reflectionless solutions exist for the problem with one-sided NZBG.
This situation is similar to what happens for the focusing and defocusing NLS equation with asymmetric NZBG \cite{PHYSD2016,JMP2014}
as well for the defocusing Manakov system with non-parallel boundary conditions \cite{EAJAM2022}.
(ii)~Since $r_\pm(\zeta,z)$ appear in the denominator of the jump matrices in the inverse problem
(cf.\ section~\ref{s:inverse}), it ensures that the jump condition does not introduce singularities in the Riemann-Hilbert problem.

\subsection{Discrete eigenvalues and residue conditions}

A discrete eigenvalue of the scattering problem is a value $\zeta\in D^+\cup D^-$
for which there exists a nontrivial solution $v(t,z,\zeta)$ to the scattering problem in \eref{e:Laxpair}
with entries in $L^2(\Real,\d t)$.
These eigenvalues occur for $\zeta\in D^+_{\mathrm{out}}$ iff the functions
$\phi(t,z,\zeta)$ and $\psi(t,z,\zeta)$
are linearly dependent (i.e., iff $a(\zeta,z)=0$); for $\zeta\in
D^-_{\mathrm{out}}$ iff the functions $\bar{\psi}(t,z,\zeta)$ and
$\bar{\phi}(t,z,\zeta)$ are linearly dependent (i.e., iff
$\bar{a}(\zeta,z)=0$);
for $\zeta\in D^-_{\mathrm{in}}$ iff the functions
$\phi(t,z,\zeta)$ and $\bar{\psi}(t,z,\zeta)$ are linearly dependent (i.e.,
iff $\bar{b}(\zeta,z)=0$); finally, for $\zeta\in D^+_{\mathrm{in}}$ iff the
functions $\psi(t,z,\zeta)$ and $\bar{\phi}(t,z,\zeta)$ are linearly dependent
(i.e., iff $b(\zeta,z)=0$).
The conjugation symmetry
\eref{symmSCAT1azvar} and the second symmetry
\eref{Symm2SCATTzvar} then imply that the discrete eigenvalues
occur in quartets: $\left\{\zeta_n\,, \zeta_n^*\,, -A^2/\zeta_n\,,
-A^2/\zeta_n^*\right\}_{n=1}^N$.

Here we assume that discrete eigenvalues are simple and finite in number, and
that there are no spectral singularities, i.e., no real zeros
of the scattering coefficients \cite{CM2005v379p21,JMP2007v48p123502,SIMA34p759,CPAM42p895}.
Establishing conditions on the asymptotic amplitudes and phases that guarantee absence
of spectral singularities is an interesting problem, but is beyond the scope of this paper.

Next, we derive the residue conditions at each of the discrete eigenvalues.
First, consider a discrete eigenvalue $\zeta_n\in D^+_{\mathrm{out}}$, i.e., a simple zero of $a(\zeta,z)$.
The Wronskian relation
\eref{Wrapaz} imply that $\psi(t,z,\zeta_n)$ and $\phi(t,z,\zeta_n)$ must be proportional:
\begin{equation}
\psi(t,z,\zeta_n)=b_n(z)\phi(t,z,\zeta_n)\,,
\label{e:b_n}
\end{equation}
for some $b_n\ne0$ independent of $t$.
In terms of the modified eigenfunctions
$\mu=(\mu^-\,\, \mu^+)=\Phi\,\e^{-ikt\sigma_3}$ and $\nu=(\nu^+\,\, \nu^-)=\Psi\,\e^{-i\lambda t\sigma_3}$ we obtain
\begin{equation}
\Res_{\zeta=\zeta_n}[\nu^+(t,z,\zeta)/a(\zeta,z)]=C_n(z)\,\e^{-i\zeta_nt}\mu^+(t,z,\zeta_n)\,,
\label{e:C_n}
\end{equation}
where $C_n(z)=b_n(z)/a'(\zeta_n,z)$ is the norming constant associated to the eigenvalue $\zeta_n\in D_+$.
Hereafter, primes will denote differentiation with respect to the spectral parameter $\zeta$.
Note that, since we assumed the eigenvalues to be simple, we have $a(\zeta_n,z)=0$ and $a'(\zeta_n,z)\ne0$.
Moreover, for all
$\zeta_n\in D^+_{\mathrm{out}}$, we have $\zeta_n^*\in D^-_{\mathrm{out}}$, and,
similarly,
$\bar{a}(\zeta_n^*,z)=0$ and $\bar{a}'(\zeta_n^*,z)\ne0$.
Correspondingly, one has
\bse
\begin{gather}
\bar{\psi}(t,z,\zeta_n^*)=\bar{b}_n(z)\bar{\phi}(t,z,\zeta_n^*)\,,
\\
\Res_{\zeta=\zeta_n^*}[\nu^-(t,z,\zeta)/\bar{a}(\zeta,z)]=\bar{C}_n(z)\,\e^{i\zeta_n^* t}\mu^-(t,z,\zeta_n^*)\,,
\label{e:Cb_n}
\end{gather}
\ese
with $\bar{C}_n(z)=\bar{b}_n(z)/\bar{a}'(\zeta_n^*,z)$.
The first symmetry implies that
$\bar{b}_n(z)=-b^*_n(z)$ and $a'(\zeta_n,z)=(\bar{a}'(\zeta_n^*,z))^*$,
so that
\begin{equation}
\bar{C}_n=-C_n^*\,.
\end{equation}
In turn, $\hat{\zeta}_n=-A^2/\zeta_n^*\in D^+_{\mathrm{in}}$, and as a result of the second symmetry $b( -A^2/\zeta_n^*,z)=0$. Then, from the Wronskian representations
it follows
\begin{equation}
\psi(t,z,-A^2/\zeta_n^*)=\hat{b}_n(z)\bar{\phi}(t,z,-A^2/\zeta_n^*)\,.
\end{equation}
From the second symmetry for the eigenfunctions we have
$(-i\zeta_n^*/q_+(z))\bar{\psi}^*(t,z,\zeta_n^*)=\hat{b}_n(z)\bar{\phi}(t,z,\zeta_n^*)$,
and on the other hand
$(-i\zeta_n^*/q_+(z))\bar{\psi}(t,z,\zeta_n^*)\equiv (-i\zeta_n^*/q_+(z))\bar{b}_n(z)\bar{\phi}(t,z,\zeta_n^*)\,$.
Therefore,
\begin{equation}
\hat{b}_n(z)=-\frac{i\zeta_n^*}{q_+(z)}\bar{b}_n(z)\,.
\end{equation}
Furthermore, $\hat{\zeta}_n^*\equiv -A^2/\zeta_n\in D^-_{\mathrm{in}}$ and from the second symmetry it follows that $\bar{b}(z,-A^2/\zeta_n)=0$ and the eigenfunctions
$\=\psi(t,z,\zeta)$ and $\phi(t,z,\zeta)$ are proportional for $\hat{\zeta}_n^*$:
\begin{gather}
\=\psi(t,z,-A^2/\zeta_n)=\bar{\hat{b}}_n(z)\phi(t,z,-A^2/\zeta_n^*)\,.
\end{gather}
Using again the second symmetry we find
\begin{equation}
\bar{\hat{b}}_n(z)=-\frac{i\zeta_n}{q_+^*(z)}b_n(z)\,.
\end{equation}
Note that the proportionality constants satisfy different symmetries for discrete eigenvalues inside and outside
$\mathcal{C}$: for discrete eigenvalues outside $\mathcal{C}$ one has $\bar{b}_n(z)=-b_n^*(z)$, while inside $\mathcal{C}$ one has
$\bar{\hat{b}}_n(z)=\hat{b}_n^*(z)$.
This is a result of the problem having asymmetric BC as $t\to\pm\infty$.

Similar symmetries exist for the norming constants associated to eigenvalues inside $\mathcal{C}$, namely:
\bse
\label{e:Chat&Cbhat}
\begin{gather}
\Res_{\zeta=\hat{\zeta}_n}[\nu^+(t,z,\zeta)/b(\zeta,z)]=\hat{C}_n(z)\,\e^{i \zeta_n^* t}\mu^-(t,z,\hat{\zeta}_n)\,,
\\
\Res_{\zeta=\hat{\zeta}_n^*}[\nu^-(t,z,\zeta)/\bar{b}(\zeta,z)]=\bar{\hat{C}}_n(z)\,\e^{-i\zeta_n t}\mu^+(t,z,\hat{\zeta}_n^*)\,,
\end{gather}
with
\begin{equation}
\hat{C}_n(z)=\frac{\hat{b}_n(z)}{b'(\hat{\zeta}_n,z)}\,,
\qquad
\bar{\hat{C}}_n(z)=\frac{\bar{\hat{b}}_n(z)}{\bar{b}'(\hat{\zeta}_n^*,z)}\,.
\end{equation}
\ese
Note $\bar{\hat{b}}_n(z)=\hat{b}_n^*(z)$ but inside $\mathcal{C}$ the first symmetry implies $\bar{b}'(\hat{\zeta}_n^*,z)=-b'(\hat{\zeta}_n,z)$,
so the norming constants for complex conjugate eigenvalues inside $\mathcal{C}$ still satisfy
\begin{equation}
\label{e:symmCh}
\bar{\hat{C}}_n(z)=-\hat{C}_n^*(z)\,.
\end{equation}
Finally, note that the definition of $\hat{C}_n(z)$ and
the second symmetry imply, respectively,
$$
\hat{C}_n(z)=  -\frac{i\zeta_n^*}{q_+(z)}\frac{\bar{b}_n(z)}{b'(\hat{\zeta}_n,z)}\,,\qquad
b'(\hat{\zeta}_n,z)=  -  i\frac{(\zeta_n^*)^3}{q_+(z)A^2}\bar{a}'(\zeta_n^*,z)  \,.
$$
We then obtain the following relationship between the norming constants inside and outside $\mathcal{C}$:
\begin{equation}
\label{e:symmCbh}
\hat{C}_n(z)=  (A/\zeta_n^*)^2\bar{C}_n(z)\equiv -(A/\zeta_n^*)^2C_n^*(z)\,.
\end{equation}

\subsection{Asymptotic behavior as $\zeta\to\infty$ and $\zeta\to0$}
\label{sec:asymptzvar}

To normalize the inverse problem in section~\ref{s:inverse}, we will need the asymptotic behavior of the
eigenfunctions and the scattering coefficients.
Note that $k\to\infty$ corresponds to $\zeta\to\infty$ in Sheet $\mathrm{I}$,
and $\zeta\to0$ in Sheet $\mathrm{II}$. Standard
Wentzel-Kramers-Brillouin (WKB) expansions in the scattering problem
rewritten in terms of $\zeta$ yield the following asymptotic behaviors for the eigenfunctions:
\bse
\label{e:asymphi&psi}
\begin{gather}
\Psi_d(t,z,\zeta)\,\e^{-i\lambda(\zeta)t\sigma_3}=I_2+o(1)\,, \quad
\Psi_o(t,z,\zeta)\,\e^{-i\lambda(\zeta)t\sigma_3}=\frac{i}{\zeta}\sigma_3Q(t,z)+o(1/\zeta)\,,\qquad
\zeta\to\infty\,,\\
\Psi_o(t,z,\zeta)\,\e^{-i\lambda(\zeta)t\sigma_3}=\frac{i}{\zeta}\sigma_3Q_+(z)+O(1)\,, \quad
\Psi_d(t,z,\zeta)\,\e^{-i\lambda(\zeta)t\sigma_3}=Q(t,z)Q_+^{-1}(z)+o(1)\,, \qquad
\zeta\to0\,,
\\
\Phi_d(t,z,\zeta)\,\e^{-ik(\zeta)t\sigma_3}=I_2+o(1)\,, \quad
\Phi_o(t,z,\zeta)\,\e^{-ik(\zeta)t\sigma_3}=\frac{i}{\zeta}\sigma_3Q(t,z)+o(1/\zeta)\,,\qquad
\zeta\to\infty\,,\\
\Phi_d(t,z,\zeta)\,\e^{-ik(\zeta)t\sigma_3}=I_2+O(\zeta)\,, \quad
\Phi_o(t,z,\zeta)\,\e^{-ik(\zeta)t\sigma_3}=-\frac{i\zeta}{A^2}\sigma_3Q(t,z)+o(\zeta)\,,\qquad
\zeta\to0\,,
\end{gather}
\ese
where subscripts ``$d$'' and ``$o$'' denote diagonal and off-diagonal part of a matrix.

Under the assumption \eref{e:H1} for the potential,
the Wronskian representations \eref{Wrapz} and
the above asymptotic behavior of the eigenfunctions then yield the
following asymptotic behavior for the scattering coefficients at large $\zeta$:
\bse
\label{e:asymscatmatrix}
\begin{gather}
\lim_{\zeta\to \infty}a(\zeta,z)
=1,\quad \zeta\in D^+_{\mathrm{out}}\cup \Real\,,
\\
\lim_{\zeta\to \infty}\bar{a}(\zeta,z)
=1,\quad \zeta\in
D^-_{\mathrm{out}}\cup \Real\,,
\\
\lim_{\zeta\to\infty} b(\zeta,z)=
\lim_{\zeta\to\infty} \bar{b}(\zeta,z)=0,\quad \zeta\in \Real\,.
\end{gather}
\ese
Using \eref{reflcoeffz} and \eref{e:reflcoeff_+} one then obtains
\be
\lim_{\zeta\to\infty}r_{\pm}(\zeta , z) = 0\,, \qquad \zeta \in \Real\,.
\ee
Similarly, the asymptotic behavior of the scattering
coefficients and the reflection coefficients as $\zeta\to0$ is as follows:
\bse
\begin{gather}
b(\zeta,z) =  \frac{iq_+^*(z)}{\zeta}+O(1),\quad \zeta\in D^+_{\mathrm{in}}\cup(-A,A)\,,\\
\bar{b}(\zeta,z) = \frac{iq_+(z)}{\zeta}+O(1),\quad \zeta\in D^-_{\mathrm{in}}\cup(-A,A)\,,
\\
\lim_{\zeta\to0}a(\zeta,z) = \lim_{\zeta\to0}\bar{a}(\zeta,z)=0,\quad \zeta\in(-A,A)\,,
\\
r_{\pm}(\zeta, z) = O(1/\zeta^2)\,,\qquad \zeta \in (-A, A)\,.
\end{gather}
\ese
Note that $b$, $\bar{b}$ and $r_{\pm}$ all have poles at $\zeta=0$.
We will see that this is not an obstacle to the formulation of the inverse problem, however.

\subsection{Behavior at the branch points}

Next we discuss the behavior of the Jost eigenfunctions and the scattering coefficients at the points $\zeta = \pm iA$,
which correspond to
the branch points $k=\pm iA$ of $\lambda(k)$ in the $k$-plane,
and are therefore still referred to as branch points even if there is no branching in the $\zeta$-plane.
Since $\lambda(\pm iA) = 0$, at $\zeta=\pm iA$ the two columns of
$Y_+(\zeta,z)$ become linearly dependent, and the two exponentials $\e^{\pm i\lambda(\zeta)t}$ reduce to the identity.
It is convenient to introduce the weighted spaces
$L^{1,j}(\Real_t^\pm):=\{f:\Real\to \Complex~ | (1+|t|)^j f(t)\in L^{1}(\Real_t^\pm)\}$.
One can
use the first part of the Lax pair~\eref{e:Laxpair} to define the Jost solutions rigorously as solutions of the following integral equations:
\bse
\label{e:intequations}
\begin{gather}
\label{e:intequationsa}
\Phi(t,z,\zeta) = \e^{ik(\zeta)t\sigma_3} + \int_{-\infty}^{t}\,\e^{ik(\zeta)(t-y)\sigma_3} Q(y,z) \Phi(y,z,\zeta) \d y,\\
\label{e:intequationsb}
\Psi(t,z,\zeta) = Y_+(\zeta,z)\,\e^{i\lambda(\zeta)t \sigma_3} - \int_t^\infty K_+(t-y,z,\zeta) \Delta Q_+(y,z) \Psi(y,z,\zeta) \d y\,,
\end{gather}
\ese
where $\Delta Q_+(y,z) = Q(y,z) -Q_+(z)$ and $K_+(y,z,\zeta):= Y_+(z,\zeta) \e^{i\lambda(\zeta)y\sigma_3} Y_+^{-1}(z,\zeta)$.
Notice that even though $Y_+(z,\pm iA)$ is not invertible, the term $K_+(t-y,z,\zeta)$ appearing in the \eref{e:intequationsb} remains finite as $\zeta \to \pm iA$:
\be
\lim_{\zeta \to \pm iA} K_+(\xi,z,\zeta)= I_2 + \xi (Q_+(z) \mp A \sigma_3)\,,
\qquad
\lim_{\zeta \to \pm iA} \partialderiv{K_+(\xi,z,\zeta)}\zeta= O_{2\times 2}\,,
\ee
with $\xi = t-y$. Thus, if $q(t,z) - q_{+}(z) \in L^{1,1}(\Real^+_t)$, the integral in \eref{e:intequationsb} is convergent at $\zeta=\pm iA$.
Moreover, $\Psi(t,z,\zeta)$ is well-defined and continuous at the branch points $\zeta = \pm iA$, with
\be
\label{e:psi1}
\Psi(t,z,\zeta) = \Psi_{\pm,1}(t,z) + o(1), \quad \zeta \to \pm iA\,,
\ee
where $\Psi_{\pm,1}(t,z)\equiv\big(\psi_{\pm,1}(t,z),\=\psi_{\pm,1}(t,z)\big):= \Psi(t,z,\pm iA)$.

Moreover, if $q(t,z) - q_+(z) \in L^{1,2}(\Real^+_t)$, it follows that  $\partial \Psi_+(t,z,\zeta)/\partial \zeta$ is well-defined and continuous as $\zeta \to \pm iA$.
Therefore one obtains
\bse
\label{e:psi2}
\be
\Psi(t,z,\zeta) = \Psi_{\pm,1}(t,z) + \Psi_{\pm,2}(t,z)(\zeta \mp iA) + o(\zeta \mp iA), \quad \zeta \to \pm iA,
\ee
with $\Psi_{\pm,1}(t,z)$ given above, and
\be
\Psi_{\pm,2}(t,z) \equiv \big(\psi_{\pm,2}(t,z),\=\psi_{\pm,2}(t,z)\big):= \left.\partialderiv{\Psi(t,z,\zeta)}\zeta\right|_{\zeta=\pm iA}.
\ee
\ese
Higher order expansions can be found similarly by placing further
restrictions on the potential and looking at higher order derivatives in $\zeta$.

Recall that $\det \Psi(t,z,\pm iA)=0$, and therefore the columns of $\Psi(t,z,\pm iA)$ are proportional to each other. Using the asymptotic behavior of $\Psi(t,z,\zeta)$ when $t \to  \infty$ and the fact that $\lambda(\zeta) = 0$ when $\zeta \to \pm iA$, one can show that
\be
\label{e:branchbehaviorofPsi}
\psi(t,z,\pm iA) = \pm e^{-i \arg q_+(z)} \=\psi(t,\pm iA).
\ee
On the other hand, the Jost eigenfunction $\Phi(t,z,\zeta)$ is not defined at $\zeta=\pm iA$, but the individual columns are continuous at the appropriate branch point (cf.\eref{e:intequationsa}). Namely, $\phi(t,z,iA)$ and $\=\phi(t,z,-iA)$ are well-defined and continuous. Using the Wronskian representations~\eref{Wrapaz} and \eref{Wrapbz}, one can obtain the following branching behavior of the scattering coefficients.

If $q(t,z) - q_+(z) \in L^{1,1}(\Real^+_t)$ then using \eref{e:psi1} one obtains
\bse
\begin{align}
    a(\zeta,z) &= \Wr (\psi_{+,1}(t,z), \phi(t,z,\zeta)) + o(1)\,, \quad
    \zeta \to iA\,,\\
\bar{a}(\zeta,z)&=\Wr(\=\phi(t,z,\zeta),\bar{\psi}_{+,1}(t,z))+o(1),\quad
    \zeta \to -iA \,,\\
b(\zeta,z)&=\Wr(\bar{\phi}(t,z,\zeta),\psi_{+,1}(t,z))+ o(1),\quad
    \zeta \to -iA\,,\\
\bar{b}(\zeta,z)&=\Wr(\bar{\psi}_{+,1}(t,z),\phi(t,z,\zeta))+o(1),\quad
    \zeta \to iA\,.
\end{align}
\ese
Similarly, assuming $q(t,z) - q_+(z) \in L^{1,2}(\Real^{+}_{t})$ and using \eref{e:psi2} we have:
\bse
\begin{align}
    a(\zeta,z) &= \Wr (\psi_{+,1}(t,z), \phi(t,z,\zeta)) + \Wr (\psi_{+,2}(t,z), \phi(t,z,\zeta)) (\zeta - iA) + o(\zeta - iA)\,, \quad
    \zeta \to iA\,,\\
\bar{a}(\zeta,z)&=\Wr(\=\phi(t,z,\zeta),\bar{\psi}_{+,1}(t,z)) + \Wr(\=\phi(t,z,\zeta),\bar{\psi}_{+,2}(t,z)) (\zeta + iA)+ o(\zeta + iA),\quad
    \zeta \to -iA \,,\\
b(\zeta,z)&=\Wr(\bar{\phi}(t,z,\zeta),\psi_{+,1}(t,z)) + \Wr(\bar{\phi}(t,z,\zeta),\psi_{+,2}(t,z)) (\zeta + iA)+ o(\zeta + iA),\quad
    \zeta \to -iA\,,\\
\bar{b}(\zeta,z)&=\Wr(\bar{\psi}_{+,1}(t,z),\phi(t,z,\zeta)) + \Wr(\bar{\psi}_{+,2}(t,z),\phi(t,z,\zeta)) (\zeta - iA)+ o(\zeta - iA),\quad
    \zeta \to iA\,.
\end{align}
\ese
One could continue this analysis by placing further restrictions on the potential if higher order terms of the scattering coefficients are needed.

Finally, we discuss the limiting behavior of the reflection coefficients near the branch points.
Recalling the definition of the reflection coefficient from the left and right,~\eref{reflcoeffz} and \eref{e:reflcoeff_+} imply that the branch points $\pm iA$ are in the domain of the latter, while the reflection coefficients from the left are only defined for $\zeta\in \Real$. To find the branch behavior of $r_+(\zeta,z)$ (resp. $\=r_+(\zeta,z)$) near $\zeta=iA$ (resp. $\zeta=-iA$), we first  compare the Wronskian representations of the scattering coefficients \eref{Wrapaz} and \eref{Wrapbz} with the proportionality relations \eref{e:branchbehaviorofPsi} and obtain
\be
    a(iA,z) = \e^{-i \arg q_+(z)} \=b(iA,z)\,,\qquad
    b(-iA,z) = -\e^{-i \arg q_+(z)} \=a(-iA,z)\,.
\ee
Thus, if $q(t,z) - q_+(z) \in L^{1,1}(\Real^+_t)$, \eref{e:reflcoeff_+} yields
\be
\lim_{\zeta \to iA} r_+(\zeta,z) = -\e^{-i \arg q_+(z)}\,,
\quad
\lim_{\zeta \to -iA} \=r_+(\zeta,z) = \e^{-i \arg q_+(z)}\,.
\ee

%%%%%%%%%%%%%%%%%%%%%%%%%%%%%%%%%%%%%%%%%%%%%%%%%%%%%%%%%%%%%%%%%%%%%%%%%%%%%%%%%%%%%%%%%%%%%%%%%%%%%%%%%%%%%%%%%%%%%
\section{Inverse problem}
\label{s:inverse}

We now discuss the inverse problem in the IST, namely the reconstruction of the solution of the MBEs~\eqref{e:MBE}
from the knowledge of the scattering data.
We
formulate the inverse problem of the IST in terms of a matrix Riemann-Hilbert problem (RHP)
for a suitable set of sectionally meromorphic functions in $D^+ \cup D^-$, with assigned jumps across $\Real \cup \mathcal{C}$, i.e., the oriented contour in the complex $\zeta$-plane as in Fig~\ref{fig:zplaneonesided}.
We show that one can obtain two different RHP formulations, namely a RHP ``from the left'' and one ``from the right'',
depending on which scattering relation is used.

\subsection{Riemann-Hilbert problem from the left}

We begin by introducing the following meromorphic matrix-valued function $M(t,z,\zeta)$
based on the analyticity properties of the Jost eigenfunctions and scattering coefficients
discussed in section~\ref{s:direct}:
\be
\label{e:M}
\displaystyle
M(t,z,\zeta)=
\begin{cases}
      \left(
      \dfrac{\psi(\zeta)}{a(\zeta)}\e^{-i\lambda(\zeta)t}\,, \,\phi(\zeta) \e^{ik(\zeta)t}
      \right),
      & \zeta\in D^+_{\text{out}}\,,
      \vspace{3mm}
      \\
      \left( \dfrac{\bar{\psi}(-\zeta^*)}{\=b(-\zeta^*)}\e^{-i\lambda(\zeta^*) t}\,, \,\phi(-\zeta^*) \e^{-ik(\zeta^*)t}\,,\,\right),
      & \zeta\in D^-_{\text{in}}\,,
      \vspace{3mm}
      \\
      \left(
      \=\phi(-\zeta^*)\e^{ik(\zeta^*) t}\,, \,\dfrac{\psi(-\zeta^*)}{b(-\zeta^*)} \e^{i\lambda(\zeta^*)t}
      \right),
      & \zeta\in D^+_{\text{in}}\,,
      \vspace{3mm}
      \\
      \left(
      \bar{\phi}(\zeta)\e^{-ik(\zeta)t}\,, \,\dfrac{\bar{\psi}(\zeta)}{\=a(\zeta)} \e^{i\lambda(\zeta)t}
      \right),
      & \zeta\in D^-_{\text{out}}\,,
\end{cases}
\ee
where the $t$ and $z$ dependence of the eigenfunctions and scattering coefficients in the right-hand side was omitted for brevity.

Using the scattering relation form the left~\eref{scatt_mat_z} as well as the second symmetry~\eref{Symm2SCATTzvar}, \eref{e:Symm2Jost} along with the fact that $-A^2/\zeta = -\zeta^*$ when $\zeta \in \mathcal{C}$, one can find the jump condition across $\zeta \in \Real \cup \mathcal{C}$ as
\bse
\label{e:Jumpcondition}
\be
M^+(t,z,\zeta) = M^-(t,z,\zeta)\,J_1(t,z,\zeta)\,,\qquad \zeta \in \Real \,,
\label{e:Jumpcondition1}
\ee
where
the superscripts $\pm$ of $M(t,z,\zeta)$ denote the limit being taken from the  left/right of the negative/positive side of the oriented contour in complex $\zeta$-plane, respectively, and where
\be
\label{e:J1def}
J_1(t,z,\zeta) =
\begin{cases}
  J_o(t,z,\zeta)\,, &  \zeta\in (-\infty , -A) \cup (A, \infty)\,,\\
  J_o(t,z,-\zeta)\,, & \zeta\in (-A,A)\,,
\end{cases}
\ee
with
\be
\displaystyle
J_o(t,z,\zeta) =
\begin{cases}
\begin{pmatrix}
     (1-r_-(z,\zeta)\,\=r_-(z,\zeta))\e^{i(k(\zeta)-\lambda(\zeta))t} & -\=r_-(z,\zeta) \e^{2ik(\zeta)t} \\
     r_-(z,\zeta) \,\e^{-2i\lambda(\zeta)t} & \e^{i(k(\zeta)-\lambda(\zeta))t}
\end{pmatrix}
\,, & \zeta\in (-\infty , -A) \cup (A, \infty)\,,
\vspace{3mm}\\
\begin{pmatrix}
    \e^{-i\zeta t}
      &  \dfrac{1}{r_-(z,\zeta)} \e^{-2i\lambda(\zeta)t}
     \vspace{1mm}
     \\
     -\dfrac{1}{\=r_-(z,\zeta)}\,\e^{-2ik(\zeta)t} & \Big(1-\dfrac{1}{r_-(z,\zeta)\,\=r_-(z,\zeta)}\Big) \e^{-i\zeta t}
\end{pmatrix}
\,, & \zeta\in (-A,A)\,.
\end{cases}
\label{e:Jodef}
\ee
\ese
Note that, as a result of having augmented the RHP to circumvent the nonlocality,
there is no jump across $\mathcal{C}$;
i.e., $M(t,z,\zeta)$ is analytic there.
Note also that $\det J_o(\zeta,z) = \e^{2i\zeta t}$, which implies $\det J_1(\zeta,z) = 1$.

Recall that the asymptotic behavior of the eigenfunctions and the scattering matrix
is given by \eref{e:asymphi&psi} and \eref{e:asymscatmatrix}.
Accordingly, the sectionally meromorphic matrices $M^\pm(t,z,\zeta)$ in~\eref{e:M} have the following asymptotic behavior:
\bse
\label{e:Masymp}
\begin{gather}
M^\pm(t,z,\zeta) = M_\infty + o(1)\,,\qquad \zeta\to \infty\,, \quad \zeta \in D_\pm^{\text{out}}\,,
\\
M^\pm(t,z,\zeta) = M_{o} + o(1)\,,\qquad \zeta\to 0\,, \quad \zeta \in D_\pm^{\text{in}}\,,
\end{gather}
\ese
where $M_\infty = M_o = I_{2\times2}$.
Moreover, one can show that the meromorphic matrices $M^\pm(t,z,\zeta)$ satisfy the following residue conditions
at the discrete eigenvalues:
\bse
\label{e:residueconditions}
\begin{gather}
\Res_{\zeta = \zeta_{n}} M(t,z,\zeta) =
  \left( ~C_n(z) \e^{-i\zeta_nt} M_2(t,z,\zeta_n)~, ~\mathbf{0}~ \right) \,,
\\
\Res_{\zeta = \zeta_{n}^*} M(t,z,\zeta) =
  \left( ~\mathbf{0}~, ~\=C_n(z) \e^{i\zeta_n^*t} M_1(t,z,\zeta_n^*)~ \right) \,,
\\
\Res_{\zeta = -\hat{\zeta}_{n}^*} M(t,z,\zeta) =
  \left( ~\mathbf{0}~, ~\hat{C}_n(z) \e^{i\zeta_n^*t} M_1(t,z,-\hat{\zeta}_n^*)~ \right) \,,
\\
\Res_{\zeta = -\hat{\zeta}_{n}} M(t,z,\zeta) =
  \left( ~\={\hat{C}}_n(z) \e^{-i\zeta_nt} M_4(t,z, -\hat{\zeta}_n)~, ~\mathbf{0}~ \right) \,,
\end{gather}
where $M_1$ and $M_2$ denote the columns of $M(t,z,\zeta)$ and the constants
$C_n, \=C_n, \hat{C}_n $ and $\={\hat{C}}_n$ were defined in~\eref{e:C_n}, \eref{e:Cb_n} and \eref{e:Chat&Cbhat}, respectively.
\ese

Summarizing, the RHP from the left consists in determining a matrix function $M(t,z,\zeta)$,
meromorphic in $\Complex\setminus \Real$,
satisfying the jump condition~\eqref{e:Jumpcondition},
the normalization condition~\eqref{e:Masymp}
and the residue conditions~\eqref{e:residueconditions}.
The minimal set of scattering data needed to define the RHP is comprised of:
(i) the reflection coefficient $r_-(0,\zeta)$ for $\zeta\in(-\infty,-A)\cup(A,\infty)$,
which (as discussed in detail in section~\ref{s:propagation})
combined with the ``boundary conditions'' $D_-(\zeta,z)$ and $P_-(\zeta,z)$
determines $r_-(z,\zeta)$ via~\eqref{e:r-propagation}
(and also $D_+(\zeta,z)$ and $P_+(\zeta,z)$ via~\eqref{e:P+})
for all $\zeta\in\Real$ and all $z>0$,
(ii) the discrete eigenvalues $\zeta_1,\dots,\zeta_N$
and (iii) the associated norming constants $C_n(z)$, $C_n^*(z)$, $\=C_n(z)$, $\hat{C}_n(z)$ and $\={\hat{C}}_n(z)$
satisfying the symmetries \eqref{e:Cb_n}, \eqref{e:symmCh} and \eqref{e:symmCbh}
(again, see section~\ref{s:propagation} for a detailed discussion of the propagation of the norming constants).

Next we show how the above RHP can be converted to a set of linear algebraic-integral equations.
We introduce the standard Cauchy projectors:
\be
\label{e:Cauchyprojector}
\displaystyle
\big(P^\pm f\big)(\zeta) = \dfrac{1}{2\pi i} \int_{\Real} \dfrac{f(s)}{s - \zeta}\d s\,, \qquad \zeta \in \Complex\setminus\Real,
\ee
which are well-defined for any function $f \in L^1(\Real)$.
If $f^\pm$ are analytic in $\Complex^\pm$ and $f^\pm = O(1/\zeta)$ as $\zeta \to \infty$ in the appropriate half-plane, then $P^\pm(f^\pm) = \pm f^\pm$ and $P^\mp(f^\pm) = 0$.
Applying the Cauchy projectors~\eref{e:Cauchyprojector} to the RHP defined by equations~\eref{e:M}, \eref{e:Jumpcondition}
and \eref{e:Masymp} yields the solution of the RHP in terms of the following system of linear algebraic-integral equations:
\bse
\label{e:SolutiontoRHP}
\begin{align}
\displaystyle
&M(t,z,\zeta) =
M_\infty +  \frac{1}{2\pi i}\int_{\Real} M^-(s)L(s)\frac{\d s}{s-\zeta}
\notag
\\
&\kern8em
+ \sum_{n=1}^N\left(
  \dfrac{\Res_{\zeta = \zeta_{n}}M(\zeta)}{\zeta-\zeta_n}
  + \dfrac{\Res_{\zeta = \zeta_{n}^*}M(\zeta)}{\zeta - \zeta_n^*}
  + \dfrac{\Res_{\zeta = -\hat{\zeta}_{n}}M(\zeta)}{\zeta+\hat{\zeta}_n}
  + \dfrac{\Res_{\zeta = -\hat{\zeta}_{n}^*}M(\zeta)}{\zeta+\hat{\zeta}_n^*}
\right)\,,
\label{e:RHPsolutionM}
\\
&
M_1(t,z,\zeta_n^*) =
\begin{pmatrix}
1\\0
\end{pmatrix}
+ \frac{1}{2\pi i}\int_{\Real} \big(M^-(s)L(s)\big)_1\,\frac{\d s}{s-\zeta_n^*}+
\sum_{n=1}^N \left(\dfrac{C_n \e^{-i\zeta_n t} M_2(\zeta_n)}{(\zeta_n^*-\zeta_n)}
+
\dfrac{\={\hat{C}}_n \e^{-i\zeta_n t} M_4(\hat{\zeta}_n^*)}{(\zeta_n^*+\hat{\zeta}_n)}
\right),
\\
&
M_2(t,z,\zeta_n) =
\begin{pmatrix}
0\\1
\end{pmatrix}
+ \frac{1}{2\pi i}\int_{\Real} \big(M^-(s)L(s)\big)_2\,\frac{\d s}{s-\zeta_n}+
\sum_{n=1}^N \left(\dfrac{\=C_n \e^{i\zeta^*_n t} M_1(\zeta^*_n)}{(\zeta_n-\zeta^*_n)}
+
\dfrac{\hat{C}_n \e^{i\zeta^*_n t} M_3(\hat{\zeta}_n)}{(\zeta_n + \hat{\zeta}_n^*)}
\right),
\end{align}
\ese
where $L(\zeta) = J_1(\zeta)- I_{2 \times 2}$,
and the $(t,z)$-dependence in the right-hand side was again omitted for brevity.

Once the solution of the above
RHP has been obtained, one can reconstruct the potential in terms of the scattering data by comparing
the resulting asymptotics of the eigenfunctions in~\eref{e:RHPsolutionM} to
\eref{e:asymphi&psi}, yielding:
\be
    q(t,z) = \frac{1}{2\pi i}\int_{\Real} \big(M^-(t,z,s)L(z,s)\big)_{12}\,\d s
    - \sum_{n=1}^N \Big(1+ \frac{iA^2 }{(\zeta_n^*)^2}\Big) \=C_n(z) \e^{i\zeta_n^*t} M_{11}(t,z,\zeta_n^*)
    \,.
\ee

\subsection{Riemann-Hilbert problem from the right}

The inverse problem of the IST can also be formulated as a RHP from the right, as we discuss next.
The reason why this could also be useful is that the MBEs are not symmetric under $t\mapsto -t$,
and in the case of one-sided NZBG, the two RHPs have different properties, depending on whether
the NZBG are given in the past or in the future.
This is reflected in the fact that, as we will see below, in the RHP from the right, the jump matrix
will be given in terms of $r_+(\zeta,z)$ [cf. \eqref{e:Jumpmatrix2} below],
whereas, in the RHP from the left, the jump matrix~\eqref{e:J1def} is given in terms of $r_-(\zeta,z)$.

Like the RHP from the left, the RHP from the right can be formulated as a local RHP for a $2\times2$ matrix
involving only the eigenfunctions evaluated at $\zeta$ and the reflection coefficients from the right on $\mathcal{C}$.
Specifically, we introduce the following $2\times2$ matrix of modified eigenfunctions:
\be
\displaystyle
M(t,z,\zeta)=
\begin{cases}
      \Big(\psi(\zeta)\e^{-i\lambda(\zeta)t},\dfrac{\phi(\zeta)}{c(\zeta)} \e^{ik(\zeta)t}\Big),
      & \zeta\in D^+_{\mathrm{out}}\,,
      \vspace{3mm}
      \\
     \Big(\dfrac{iq_+^*\zeta^*}{A^2}\=\psi(-\zeta^*)\e^{-i\lambda(\zeta^*) t}\,,\,
     \dfrac{iq_+\zeta^*}{A^2}\dfrac{\phi(-\zeta^*)}{d(-\zeta^*)} \e^{-ik(\zeta^*)t}\Big)\,,
     & \zeta\in D^-_{\mathrm{in}},
      \vspace{3mm}
      \\
      \Big(\dfrac{iq_+^*\zeta^*}{A^2}\dfrac{\=\phi(-\zeta^*)}{\=d(-\zeta^*)}\e^{ik(\zeta^*)t}\,,\,
      \dfrac{iq_+\zeta^*}{A^2}\psi(-\zeta^*) \e^{i\lambda(\zeta^*)t} \Big),
      & \zeta\in D^+_{\mathrm{in}}\,,
      \vspace{3mm}
      \\
      \Big(\dfrac{\=\phi(\zeta)}{\=c(\zeta)}\e^{-ik(\zeta)t}\,,\,\=\psi(\zeta) \e^{i\lambda(\zeta)t} \Big),
      & \zeta\in D^-_{\mathrm{out}}\,,
   \end{cases}
\ee
which should be compared with \eqref{e:M}, and where the $t$ and $z$ dependence in the right-hand side
was again omitted for brevity.
Now using the scattering relation from the right~\eref{scatt_mat_z2} and the second symmetry~\eref{e:Symm2Jost},
along with the fact that $-A^2/\zeta = -\zeta^*$ when $\zeta \in \mathcal{C}$,
we obtain that the jump condition across $\zeta \in \Real$ is still expressed by \eqref{e:Jumpcondition},
with $J_1(z,\zeta)$ still given by~\eqref{e:J1def}, except that now
\be
\label{e:Jumpmatrix2}
\displaystyle
J_o(t,z,\zeta) =
\begin{cases}
\begin{pmatrix}
     \e^{i(k(\zeta)-\lambda(\zeta))t} & r_+(\zeta,z) \e^{2ik(\zeta)t} \\
     -\=r_+(\zeta,z) \,\e^{-2i\lambda(\zeta)t} & (1-r_+(\zeta,z)\,\=r_+(\zeta,z))\e^{i(k(\zeta)-\lambda(\zeta))t}
\end{pmatrix}
\,, & \zeta\in (-\infty , -A) \cup (A, \infty)\,,
\vspace{3mm}\\
\begin{pmatrix}
      \e^{i\zeta t} &  \dfrac{q_+(z)}{q_+^*(z)\,r_+(\zeta,z)} \e^{2ik(\zeta)t}
      \vspace{1mm}
      \\
     -\dfrac{q_+^*(z)}{q_+(z)\,\=r_+(\zeta,z)}\,\e^{2i\lambda(\zeta)t} & \Big(1-\dfrac{1}{r_+(\zeta,z)\,\=r_+(\zeta,z)}\Big)\e^{i\zeta t}
\end{pmatrix}
\,, & \zeta\in (-A,A)\,.
\end{cases}
\ee
Again, note that no jump is present across $\mathcal{C}$. As before, the above jump condition must be supplemented by appropriate normalization condition and residue conditions.
Since these are obtained using similar methods as above, we omit the details for brevity.

%%%%%%%%%%%%%%%%%%%%%%%%%%%%%%%%%%%%%%%%%%%%%%%%%%%%%%%%%%%%%%%%%%%%%%%%%%%%%%%%%%%%%%%%%%%%%%%%%%%%%%%%%%%%%%%%%%%%%
\subsection{Trace formulae}
\label{s:trace}

As usual, one can also obtain ``trace formulae'' to recover the analytic scattering coefficients in terms of scattering data.
We begin by defining following functions:
\bse
\begin{gather}
    \beta^+(\zeta,z) = a(\zeta,z) \prod_{n=1}^N \frac{\zeta-\zeta_n^*}{\zeta-\zeta_n}\,,
    \quad
    \beta^-(\zeta,z) = \=a(\zeta,z) \prod_{n=1}^N \frac{\zeta-\zeta_n}{\zeta-\zeta_n^*}\,,
    \\
    \alpha^+(\zeta,z)  = -\frac{i\zeta}{q_+^*(z)}\,b(\zeta,z) \prod_{n=1}^N \frac{\zeta_n(\zeta-\hat{\zeta}_n^*)}{\zeta_n^*(\zeta-\hat{\zeta}_n)}\,,
    \quad
    \alpha^-(\zeta,z)  = -\frac{i\zeta}{q_+(z)}\,\=b(\zeta,z) \prod_{n=1}^N \frac{\zeta_n^*(\zeta-\hat{\zeta}_n)}{\zeta_n(\zeta-\hat{\zeta}_n^*)}\,.
\end{gather}
\ese
Recalling \eref{Wrapz}, one can see that $\beta^\pm$ are analytic in $D^\pm_{\mathrm{out}}$, while $\alpha^\pm$ analytic in $D^\pm_{\mathrm{in}}$. Moreover, using the asymptotic behavior of the scattering coefficients~\eref{e:asymscatmatrix} one can show that
\be
\lim_{\zeta \to \infty} \beta^\pm = 1\,, \quad \zeta \in D^\pm_{\mathrm{out}} \,,
\qquad
\lim_{\zeta \to 0} \alpha^\pm = 1\,, \quad \zeta \in D^\pm_{\mathrm{in}} \,.
\ee
Also, by construction $\beta^\pm$ and $\alpha^\pm$ have no zeros.
Now we define the following sectionally analytic vector function:
\be
\label{e:N}
\displaystyle
N(t,z,\zeta)=
\begin{cases}
      \Big(\log \big( \beta^+(\zeta,z)\big)~,~ \log \big( \beta^+(-\zeta^*,z)\big)\Big)  \,, & \zeta\in D^+_{\mathrm{out}}\,,
      \vspace{3mm}
      \\
     \Big(\log \big( \alpha^-(-\zeta^*,z)\big)~,~ \log \big( \alpha^-(\zeta,z)\big)\Big)  \,, & \zeta\in D^-_{\mathrm{in}}\,,
      \vspace{3mm}
      \\
      \Big(-\log \big( \alpha^+(-\zeta^*,z)\big)~,~ -\log \big( \alpha^+(\zeta,z)\big)\Big)  \,, & \zeta\in D^+_{\mathrm{in}}\,,
      \vspace{3mm}
      \\
      \Big(-\log \big( \beta^-(\zeta,z)\big)~,~ -\log \big( \beta^-(-\zeta^*,z)\big)\Big)  \,, & \zeta\in D^-_{\mathrm{out}}\,.
   \end{cases}
\ee
Using the determinant of the scattering matrix form the left~\eref{scatt_mat_z} and \eref{e:detS} as well as the second symmetry of the scattering coefficients~\eref{Symm2SCATTzvar}, along with the fact that $-A^2/\zeta = -\zeta^*$ when $\zeta \in \mathcal{C}$, one can find the jump condition across
$\zeta \in \Real$ as
\be
\label{e:scattering:Jumpcondition}
N^+(t,z,\zeta) - N^-(t,z,\zeta) = K(z,\zeta)\,,\qquad \zeta \in \Real.
\ee
where the superscripts $\pm$ of $N(t,z,\zeta)$ denote the limit being taken from the  left/right of the negative/positive side of the oriented contour in complex $\zeta$-plane, respectively. Moreover,
\bse
\label{e:K}
\be
\displaystyle
K(z,\zeta) =
\begin{cases}
  \Big(  K_o(z,\zeta) ~,~ K_o(z,-\zeta) \Big)\,, & \zeta\in (-\infty , -A) \cup (A, \infty)\,,
\vspace{3mm}\\
   \Big(  K_o(z,-\zeta) ~,~ K_o(z,\zeta) \Big)\,, & \zeta\in (-A,A)\,,
\end{cases}
\ee
where
\be
\label{e:Ko}
\displaystyle
K_o(z,\zeta) =
\begin{cases}
 -\log \Big[ \frac{\zeta^2}{\zeta^2 + A^2} (1+ |r_-(z,\zeta)|^2) \Big]
\,, & \zeta\in (-\infty , -A) \cup (A, \infty)\,,
\vspace{3mm}
\\
\log \Big[ \frac{A^2}{\zeta^2 + A^2} \Big(1+ \frac{1}{|r_-(z,\zeta)|^2}\Big)\Big]
\,, & \zeta\in (-A,A)\,.
\end{cases}
\ee
\ese
Applying the Cauchy projectors~\eref{e:Cauchyprojector} to the RHP defined by equations~\eref{e:N} and \eref{e:scattering:Jumpcondition} yields
\be
N(t,z,\zeta) = \frac{1}{2\pi i} \int_{\Real} \frac{K(z,s)}{s-\zeta}\d s \,,
\qquad
\zeta \in \Complex \setminus \Real\,.
\ee
Now using the solution of the RHP, one can recover the analytic scattering coefficient from the knowledge of the reflection coefficients and discrete eigenvalues, as follows:
\bse
\begin{gather}
\label{e:Trace}
a(\zeta,z) = \prod_{n=1}^{N}\frac{\zeta-\zeta_n}{\zeta-\zeta_n^*} \exp \Big\{\frac{1}{2\pi i} \int_{\Real} \frac{K_1(z,s)}{s-\zeta}\d s \ \Big\}\,,
\quad \zeta \in D^+_{\text{out}}\,,
\\
\=a(\zeta,z) = \prod_{n=1}^{N}\frac{\zeta-\zeta_n^*}{\zeta-\zeta_n} \exp \Big\{-\frac{1}{2\pi i} \int_{\Real} \frac{K_1(z,s)}{s-\zeta}\d s \ \Big\}\,,
\quad \zeta \in D^-_{\text{out}}\,,
\\
b(\zeta,z) = \frac{iq_+^*(z)}{\zeta}\prod_{n=1}^{N}\frac{\zeta_n^*(\zeta-\hat{\zeta}_n)}{\zeta_n(\zeta-\hat{\zeta}_n^*)} \exp \Big\{-\frac{1}{2\pi i} \int_{\Real} \frac{K_2(z,s)}{s-\zeta}\d s \ \Big\}\,,
\quad \zeta \in D^+_{\text{in}}\,,
\\
\=b(\zeta,z) = \frac{iq_+(z)}{\zeta}\prod_{n=1}^{N}\frac{\zeta_n(\zeta-\hat{\zeta}_n^*)}{\zeta_n^*(\zeta-\hat{\zeta}_n)} \exp \Big\{\frac{1}{2\pi i} \int_{\Real} \frac{K_2(z,s)}{s-\zeta}\d s \ \Big\}\,,
\quad \zeta \in D^-_{\text{in}}\,,
\end{gather}
\ese
where $K(z,\zeta)$ is given by \eref{e:K} and the subscript $j=1,2$ denotes its $j$-th column.

We reiterate that, unlike what happens in the IST for the focusing NLS equation and for the MBE with zero background,
the above trace formulae are needed in order to obtain the correct propagation equation for the norming constants
(cf.\ section~\ref{ss:normingpropagation}).

%%%%%%%%%%%%%%%%%%%%%%%%%%%%%%%%%%%%%%%%%%%%%%%%%%%%%%%%%%%%%%%%%%%%%%%%%%%%%%%%%%%%%%%%%%%%%%%%%%%%%%%%%%%%%%%%%%%%%
\section{Asymptotics of the density matrix as $t\to\pm\infty$}
\label{s:asympdensity}

If $v(t,z,\zeta)$ is any fundamental matrix solution of the scattering problem,
$v^{-1}(t,z,\zeta)\rho(t,z,\zeta)v(t,z,\zeta)$ is time-independent.
As a consequence, evaluating the limits as $t\to \pm \infty$ one finds for $\zeta \in \Real$
\begin{gather}
\label{def:rhopm}
\rho_-(\zeta,z)=\Phi^{-1}(t,z,\zeta)\rho(t,z,\zeta)\Phi(t,z,\zeta)\,, \qquad \rho_+(\zeta,z)=\Psi^{-1}(t,z,\zeta)\rho(t,z,\zeta)\Psi(t,z,\zeta)\,,
\end{gather}
and conversely
\begin{gather}
\label{e:rho}
\rho(t,z,\zeta)=\Phi(t,z,\zeta)\rho_-(\zeta,z)\Phi^{-1}(t,z,\zeta)\equiv \Psi(t,z,\zeta)\rho_+(\zeta,z)\Psi^{-1}(t,z,\zeta)\,.
\end{gather}
Taking into account the asymptotics as $t\to \pm \infty$ of the Jost eigenfunctions we obtain
\begin{gather}
\label{e:rhopm}
\rho_-(\zeta,z)=\lim_{t\to -\infty}\e^{-ikt\sigma_3}\rho(t,z,\zeta)\e^{ikt\sigma_3}\,,
\quad
\rho_+(\zeta,z)=\lim_{t\to +\infty}\e^{-i\lambda t\sigma_3}Y_+^{-1}(\zeta,z)\rho(t,z,\zeta)Y_+(\zeta, z) \e^{i\lambda t\sigma_3}\,.
\end{gather}
Note, however, that the scattering matrix does not have a limit per se, and $\rho_\pm(\zeta,z)$ are not simply the limits
of $\rho(t,z,\zeta)$ as $t\to \pm \infty$. One can also check by direct calculation that $S\rho_+=\rho_-S=\Phi^{-1}\rho\Psi$ and
therefore
\begin{equation}
\label{e:relrho+rho-}
\rho_+(z,\zeta)=S^{-1}(\zeta,z)\rho_-(\zeta,z)S(\zeta,z)\,, \qquad \forall \zeta \in \Real\,.
\end{equation}
The above equation relates the asymptotic values of the density matrix as $t\to \pm \infty$, and allows one to obtain $\rho_+$ from knowledge of $\rho_-$ and $S$ [which in turn can be completed determined by $q(t,z)$]. Thus, one can only choose one between $\rho_\pm$, and due to causality, it makes sense to choose $\rho_-$.
Note also that the density matrix is a single-valued function of $k$, which means that $\rho(t,z,-A^2/\zeta)=\rho(t,z,\zeta)$, and the same holds for $\rho_-(\zeta,z)$
(which depends on the Jost eigenfunction $\Phi$), but not for $\rho_+(\zeta,z)$, which is defined via $\Psi$.

The properties of the density matrix $\tr \rho=0$ and $\det \rho=-1$ imply that $\tr \rho_\pm=0$ and $\det \rho_\pm =-1$. Also, since $\rho^\dagger =\rho$ for $\zeta \in \Real$, the same holds for $\rho_\pm=\rho_\pm^\dagger$. Thus we can denote the entries of $\rho_\pm$ as
\begin{equation}
\label{e:matrixrhopm}
\rho_\pm(\zeta,z)=\begin{pmatrix}
D_\pm & P_\pm \\
P_\pm^* & -D_\pm
\end{pmatrix} \qquad \zeta \in \Real\,,
\end{equation}
with $P_\pm(\zeta,z)=P_\pm(\zeta^*,z)$ . As for the density matrix itself, $D_\pm$ and $P_\pm$ are not limits of $D$ and $P$ as $t\to \pm \infty$,
as such limits in general do not exist. Combining the second symmetry of the eigenfunctions with the definition \eref{e:rhopm} we have
for all $\zeta \in \Real$:
\begin{equation}
\rho_-(-A^2/\zeta,z)=\rho_-(\zeta,z)\,, \qquad \rho_+(-A^2/\zeta,z)=\sigma_3 Q_+(z_)\rho_+(\zeta,z)Q_+^{-1}(z)\sigma_3\,.
\end{equation}
This implies that one does not have the freedom to pick the asymptotic states $\rho_\pm$ for all values of $k$ (or $\zeta$). One can only pick $\rho_\pm$ for $k\in \Real$ on first sheet, or, equivalently, for $\zeta \in (-\infty,-A]\cap [A,\infty)$. Specifically,
\bse
\label{e:D_pm&P_pmsheet1&2}
\begin{gather}
D_\pm \in \Real, \quad P_\pm \in \Complex, \quad D_\pm^2+|P_\pm|^2=1\,, \\
D_-(-A^2/\zeta,z)=D_-(\zeta,z)\,, \qquad D_+(-A^2/\zeta,z)=-D_+(\zeta,z)\,, \\
P_-(-A^2/\zeta,z)=P_-(\zeta,z)\,, \qquad P_+(-A^2/\zeta,z)=(q_+(z)/q_+^*(z)) P_+^*(\zeta,z)\,.
\end{gather}
\ese
Equation~\eref{e:relrho+rho-} can then be used to obtain $D_+$ and $P_+$ from $D_-$ and $P_-$, respectively.
Note that in principle $D_\pm$ and $P_\pm$ depend
on $k$ on each sheet.

Equations~\eref{e:rhopm} yield an explicit relation between $\rho_\pm$ and $\rho$, which, in component form, is
\bse
\begin{gather}
D_+(z,\zeta) = \lim_{t\to +\infty}\left[\frac{k}{\lambda}D-\frac{1}{\lambda}\Im (q_+^*P) \right]\,, \\
P_+(z,\zeta) = \frac{1}{2\lambda}\lim_{t\to +\infty}\e^{-2i\lambda t}\left[2iq_+D+\zeta P+\frac{q_+^2}{\zeta}P^* \right]\,.
\end{gather}
\ese
Conversely, \eref{e:rho} also implies that
\bse
\begin{gather}
\rho(t,z,\zeta) = \e^{ikt\sigma_3}\rho_-(\zeta,z)\e^{-ikt\sigma_3}+o(1),\qquad t\to -\infty\,,\\
\rho(t,z,\zeta) = Y_+(\zeta,z)\e^{i\lambda t\sigma_3}\rho_+(\zeta,z)\e^{-i\lambda t\sigma_3}Y_+^{-1}(\zeta,z)+o(1),\qquad \to -\infty,\
\end{gather}
\ese
i.e.
\bse
\label{e:DPasymptinfty}
\begin{gather}
D(t,z,\zeta) = D_- +o(1), \quad
P(t, z,\zeta) = \e^{-2ikt}P_-+o(1)\,,\qquad
t\to-\infty, \\
D(t, z,\zeta) = \frac{k}{\lambda}D_+-\frac{1}{\lambda}\Im\left(\e^{-2i\lambda t}P_+^*q_+ \right)+o(1),\qquad
t\to\infty\,,
\\
P(t, z,\zeta) = -\frac{i}{\lambda}q_+D_++\frac{\zeta}{2\lambda}\e^{2i\lambda t}P_++\frac{q_+}{2\lambda \zeta}\e^{-2i\lambda t}P_+^*+o(1),\qquad
t\to\infty\,.
\end{gather}
\ese
Note that, as in the case of ZBG and of symmetric NZBG, $P$ does not have a limit as $t\to \pm \infty$, but instead it
oscillates in time. The quantity $D$, on the other hand, has a constant limit $D_-$ as $t\to -\infty$, but due to the nonzero background radiation as $t\to +\infty$, it  also does not have a limit and instead oscillates. Moreover, $P\ne 0$ as $t\to + \infty$ even in the particular case in which $D$ and $P$ are time-independent and do tend to a limit as $t\to +\infty$, which is when $P_+=0$.
The nonzero contribution arises from the polarization induced by the limiting value $q_+$ of the optical field via~$D_+$. [Recall that the normalization $\det \rho_-=-1$ implies the constraint $D_\pm^2+|P_\pm|^2=1$, which in turn implies that one does not have the freedom to assign $D_\pm$ and $P_\pm$ independently.
In particular, in the special case when $\rho_-$ is diagonal, then $D_-=\pm 1$ and $P_-=0$].

Importantly, recall that the density matrix $\rho(t,z,k)$ describes the physical properties of the medium,
and therefore its value is independent of the choice of sign for $\lambda$ and is therefore the same
for  $\zeta\in(-\infty,-A]\cup[A,\infty)$ (i.e., the continuous spectrum on sheet I) or $\zeta\in[-A,A]$
(i.e., the continuous spectrum on sheet II).
The same is true for the first set of Jost solutions, $\Phi(t,z,\zeta)$.
Conversely, the Jost solutions $\Psi(t,z,\zeta)$ are defined explicitly in terms of $\lambda$,
and therefore take on different values
for $\zeta\in(-\infty,-A]\cup[A,\infty)$ or $\zeta\in[-A,A]$
(as discussed in section~\ref{s:symmetries}).
Similarly, $\rho_-(\zeta,z)$, which is defined in terms of $\Phi(t,z,\zeta)$, is independent of the choice of sign for $\lambda$,
whereas $\rho_+(\zeta,z)$, which is defined in terms of $\Psi(t,z,\zeta)$, depends on the sign of $\lambda$,
and is therefore different on different sheets.
Equations~\eqref{e:D_pm&P_pmsheet1&2} reflect this difference.
In turn, this sheet dependence is also reflected in~\eqref{e:DPasymptinfty},
and it should be taken into account when reconstructing the asymptotic behavior of the medium.
For example, if $P_\pm=0$, the sign of $D_+$ coincides with that of $D(z,\zeta)$ on the first sheet,
not the second one.

%%%%%%%%%%%%%%%%%%%%%%%%%%%%%%%%%%%%%%%%%%%%%%%%%%%%%%%%%%%%%%%%%%%%%%%%%%%%%%%%%%%%%%%%%%%%%%%%%%%%%%%%%%%%%%%%%%%%%
\section{Propagation}
\label{s:propagation}

Recall that in the MBE the role of the evolution variable is played by the physical propagation distance $z$, and therefore we will refer to the $z$-dependence as the propagation.
The evolution of the scattering data is where the IST formalism for the MBE differs significantly from
(and is considerably more complicated than) that for the NLS equation.

%%%%%%%%%%%%%%%%%%%%%%%%%%%%%%%%%%%%%%%%%%%%%%%%%%%%%%%%%%%%%%%%%%%%%%%%%%%%%%%%%%%%%%%%%%%%%%%%%%%%%%%%%%%%%%%%%%%%%
\subsection{Propagation of the background}

Let us first discuss the $z$-dependence of the asymptotic values of the optical field. The propagation of these values is given by the limits as $t\to \pm \infty$
of \eref{e:MBE}, i.e.
$$
\frac{\partial Q_\pm}{\partial z}=-\frac{1}{2}\lim_{t\to \pm \infty}\int [\sigma_3, \rho(t,z,k)]g(k)dk\,.
$$
Using \eref{e:rho} we can express
\begin{gather}
\rho(t,z,\zeta)=\e^{ikt\sigma_3}\rho_-(z,\zeta)\e^{-ikt\sigma_3}+o(1) \qquad \text{as } t\to -\infty\,,\\
\rho(t,z,\zeta)=Y_+(\zeta,z) \e^{i\lambda t\sigma_3}\rho_+(z,\zeta)\e^{-i\lambda t\sigma_3}Y_+^{-1}(\zeta,z)+o(1) \qquad \text{as } t\to +\infty\,.
\end{gather}
Consequently, one has:
\begin{gather}
\label{e:dzQ}
\frac{\partial Q_-}{\partial z}=0\,, \qquad \frac{\partial Q_+}{\partial z}=i w_+ [\sigma_3,Q_+]\,,
\end{gather}
where
\begin{equation}
w_+(z)=\frac{1}{2}\int\frac{ D_+(\xi,z)g(\xi)}{\lambda(\xi)} \d\xi\,.
\label{e:w_+}
\end{equation}
Recall that $\tr \rho_\pm=\tr \rho=0$ and $D_+$ has opposite sign on sheets I and II.
Since $\lambda$ also changes sign on opposite sheets, $w_+$ is single-valued, as it should be. Integrating \eqref{e:dzQ} we find
\begin{equation}
\label{e:Q_+(z)}
Q_-(z)=Q_-(0)\equiv 0\,, \qquad Q_+(z)=\e^{iW_+(z)\sigma_3}Q_+(0)\e^{-iW_+(z)\sigma_3}\,, \qquad W_+(z)=\int_0^zw_+(z')dz'\,,
\end{equation}
or simply
\be
q_-(z)=0\,, \qquad q_+(z)=\e^{2iW_+(z)}q_+(0)\,.
\label{e:qpmz}
\ee
Equations~\eqref{e:qpmz} provide the explicit $z$-dependence in the symmetries in section~\ref{s:symmetries}.
In particular, the propagation of the asymptotic eigenvector matrix $Y_+(\zeta,z)$ is given by:
\begin{gather}
Y_+(\zeta,z)=I+(i/\zeta)\sigma_3Q_+(z)=\e^{iW_+(z)\sigma_3}Y_+(\zeta,0)\e^{-iW_+(z)\sigma_3}\,,
\end{gather}
and consequently the asymptotic behavior of the Jost solutions is
\bse
\label{e:Jostpropagation}
\begin{gather}
\Phi(t,z,\zeta)=I_2\e^{ik(\zeta)t\sigma_3}(1 +o(1))\,, \qquad \text{as } t\to -\infty \\
\Psi(t,z,\zeta)=\e^{iW_+(z)\sigma_3}Y_+(\zeta,0)\e^{i(\lambda(\zeta)t-W_+(z))\sigma_3}(1 +o(1))\,, \qquad \text{as } t\to +\infty \,.
\end{gather}
\ese
We can use \eref{e:relrho+rho-} to express $2w_+(z)\sigma_3=\int (\rho_{+,d}(k,z)g(k)/\lambda)dk$ in terms of $\rho_{-,d}$ as:
\be
\label{e:rhodiagonal}
\rho_{+,d}=(S^{-1}\rho_-S)_d=\rho_{-,d}+\frac{\zeta}{2\lambda}\left( 2b\bar{b}\, D_-+P_-\bar{a}b-P_-^*a\bar{b}\right)\sigma_3\,.
\ee
Note that the above expression has an extra term proportional to $b\bar{b}$ compared to \cite{BGKL},
and this extra term is also present in the case of symmetric NZBG studied there.
Considering  \eref{e:rhodiagonal} in component form and using the symmetries \eref{symmJost1zvar} one obtains
\bse
\label{e:DP+}
\be
D_+ = \frac{\zeta}{2\lambda}\Big( \big(|a|^2 - |b|^2\big)D_- + a^*b P_- + ab^* P_-^*\Big)\,.
\ee
We can also use \eref{e:relrho+rho-} to express the off-diagonal entries of $\rho_{+}$ in terms of those of $\rho_-$:
\be
\label{e:P+}
P_+=\frac{\zeta}{2\lambda}\Big( (a^*)^2P_--(b^2P_-)^* - 2a^*b^* D_-\Big)\,.
\ee
\ese
Using \eref{reflcoeffz} and \eref{e:a2-b2} one could rewrite above $D_+$ and $P_+$ explicitly as follows:
\be
D_+ = \frac{1}{1+ |r_-|^2}\Big( \big(1 - |r_-|^2\big)D_- + 2 \Re (r_-\,P_-)\Big)\,,
\qquad
P_+=\frac{\e^{-2i\arg(a)}}{1+|r_-|^2}\Big( P_--(r_-^2\,P_-)^* - 2 r_-^* \,D_-\Big)\,,
\label{e:D+P+}
\ee
where we omitted the dependence on $\zeta$ and $z$ for brevity.
Equations~\eqref{e:D+P+}, which are the same as in the case of ZBG except for the additional presence of the factor
$\zeta/(2\lambda)$
(which is replaced by 1 in the case of ZBG and is a direct consequence of the fact that $\det S\ne1$),
show that even if the medium is initially prepared so that $P_-=0$ and $D_-=\pm 1$,
one has $P_+\ne0$ in general, since $r_-$ cannot be chosen to be identically zero.
(In the case of symmetric NZBG, $P_-=0$ does not imply $P_+=0$ except in the reflectionless case.)
The limiting values of these quantities as $z\to\infty$ will be discussed in section~\ref{s:lta}.

%%%%%%%%%%%%%%%%%%%%%%%%%%%%%%%%%%%%%%%%%%%%%%%%%%%%%%%%%%%%%%%%%%%%%%%%%%%%%%%%%%%%%%%%%%%%%%%%%%%%%%%%%%%%%%%%%%%%%
\subsection{Simultaneous solutions of the Lax pair}

To obtain propagation equations for the scattering data,
which will be done in section~\ref{ss:propagationreflection} and~\ref{ss:normingpropagation},
one needs to introduce simultaneous solutions of both parts of the Lax pair.
Since the asymptotic behavior of the Jost solutions $\Phi(t,z,\zeta)$ and $\Psi(t,z,\zeta)$ as $t\to \pm \infty$ is independent of $z$, in general they will not be solutions of \eref{e:Laxpair2}.
However, both $\Phi$ and $\Psi$ are fundamental matrix solutions of the scattering problem,
and any other solution $\Xi(t,z,\zeta)$ can then be
written as
\vspace*{-1ex}
\begin{gather}
\label{e:C_pm}
\Xi(t,z,\zeta)=\Psi(t,z,\zeta)C_+(\zeta,z)=\Phi(t,z,\zeta)C_-(\zeta,z)\,, \qquad \zeta \in \Real\,,
\end{gather}
where $C_\pm(\zeta,z)$ are $2\times 2$ matrices independent of $t$.
Then, if $\Xi(t,z,\zeta)$ is a simultaneous solution of both parts of the Lax pair \eref{e:Laxpair},
(i.e., it also satisfies $\Xi_z=T\Xi$), it follows that
\vspace*{0ex}
\begin{gather}
\label{e:Cpm_z}
\frac{\partial C_\pm}{\partial z}=\frac{i}{2}R_\pm C_\pm\,,
\end{gather}
with
\vspace*{-1ex}
\begin{gather}
\label{e:Rpm}
R_-(\zeta,z) = -2i\Phi^{-1}[T\Phi-\partial_z\Phi]\,,
\qquad
R_+(\zeta,z) = - 2i\Psi^{-1}[T\Psi-\partial_z\Psi]\,,
\end{gather}
where we have taken into account that for our one-sided NZBG $\Phi$ can be chosen independent of $z$.
Although it is not obvious a priori that the RHS of \eref{e:Rpm} is independent of $t$, \eref{e:Cpm_z} shows that it must be.
Moreover, even though $g(k)$ and $\rho(t,z,k)$ are only defined for $k\in \Real$, $R_\pm$ in \eref{e:Rpm} can be evaluated for all $\zeta\in \Real$.
In general, both $C_\pm$ and $R_\pm$ have different values on either sheet of the Riemann surface.
Using \eref{e:rho}, we can write the second operator in the Lax pair as
\vspace*{-1ex}
\begin{gather}
T(t,z,\zeta) = \frac{i\pi}{2}\mathcal{H}_k[\Phi(t,z,\zeta(\xi))\rho_-(\xi,z)\Phi^{-1}(t,z,\zeta(\xi)\,g(\xi)]
\\
= \frac{i\pi}{2}\mathcal{H}_k[\Psi(t,z,\zeta(\xi))\rho_+(\xi,z)\Psi^{-1}(t,z,\zeta(\xi)\,g(\xi)]\,,
\end{gather}
where the subscript $k$ in the Hilbert transform is to be intended as $k(\zeta)$ henceforth.
As in \cite{BGKL}, even though the individual terms on the RHS of the second equality are only single-valued for $k\in \Real$, the whole RHS
is a single-valued function for $k\in \Complex$. In fact, one can use the symmetries of the Jost solution $\Psi$ and of $\rho_+$ to show
that $T(t,z,\zeta)=T(t,z,-A^2/\zeta)$.

Next, we can evaluate the RHS of the first of \eref{e:Rpm} in the limit $t\to - \infty$.
Assuming that $\partial_z$ and the limit $t \to -\infty$ commute, we obtain
\begin{gather}
R_-(\zeta,z) =
-2i\lim_{t\to -\infty}\e^{-ikt\sigma_3}
\left[T(t,z,\zeta)\e^{ikt\sigma_3}- \partial_z \e^{ikt\sigma_3} \right]
\notag
\\ = \pi \,\mathcal{H}_k\left[\e^{-i(k-\xi)t\sigma_3} \rho_-(\xi,z)\e^{i(k-\xi)t\sigma_3}\,g(\xi)\right]\,.
\end{gather}
Now the second of \eref{e:Rpm} in the limit $t\to +\infty$ yields:
\begin{gather}
R_+(\zeta,z)= \lim_{t\to +\infty}
\Big\{ 2i\Psi^{-1}(t,z,\zeta)\partial_z \Psi(t,z,\zeta) \notag \\
\left. +\pi \e^{-i\lambda(k) t\sigma_3}Y_+(k,z)\mathcal{H}_k\left[Y_+(\xi,z)\e^{i\lambda(\xi)t\sigma_3}\rho_+(\xi,z)\e^{-i\lambda(\xi)t\sigma_3}
Y_+^{-1}(\xi,z)g(\xi)
\right]Y_+(k,z)\e^{i\lambda(k) t\sigma_3} \right\}\,.
\end{gather}
The limits can be computed explicitly by noting that
\begin{gather}
\label{e:relationforRpm}
\lim_{t\to \pm \infty}\dashint \e^{2i(k-\xi)t}\frac{f(\xi,k)}{\xi-k}\d\xi=\mp i\pi f(k,k) \qquad k\in \Real \\
\lim_{t\to \pm \infty}\dashint \e^{\pm i(\lambda(\xi)-\lambda(k))t}\frac{f(\xi,k)}{\xi-k}\d\xi
=\left\{
\begin{array}{ll}
\pm i \sigma \pi f(k,k) & k\in \Real \\
0 & k\in i[-A,A]
\end{array}
\right.
\end{gather}
where $\sigma=\pm1$ when $k$ is on sheet I or II, respectively.
Then, if $R_\pm=R_{\pm,d}+R_{\pm,o}$ where as before the subscripts
``$d$'' and ``$o$'' denote the diagonal and off-diagonal parts of the corresponding matrices, we find:
\bse
\label{e:Rpmo,d}
\begin{gather}
\label{e:Rpmd}
R_{-,d}(\zeta,z)=\pi \mathcal{H}_k[\rho_{-,d}(\xi,z)g(\xi)]\,, \qquad R_{+,d}(\zeta,z)=\pi \lambda(\zeta) \mathcal{H}_k[\rho_{+,d}(\xi,z)g(\xi)/\lambda(\xi)]
+2w_+(z)\sigma_3
\qquad \zeta\in \Complex
\\
\label{e:Rpmo}
R_{-,o}(\zeta,z)=i\pi g(k(\zeta))\rho_{-,o}(\zeta,z)\sigma_3\,,
\qquad R_{+,o}(\zeta,z)=\left\{
\begin{array}{ll}
-i\sigma \pi g(k(\zeta))\rho_{+,o}(\zeta,z)\sigma_3 & \zeta \in \Real
\\
0 & \zeta \in \mathcal{C}\setminus\left\{\pm A
\right\}
\end{array}
\right.
\end{gather}
\ese
where $\sigma=1$ for $\zeta\in (-\infty,-A]\cup [A,+\infty)$ and $\sigma=-1$ for $\zeta\in (-A,A)$.
In component form, the above equations are
\bse
\label{e:Rpmcomponents}
\begin{gather}
R_{+,11}=- R_{+,22}=\dashint\frac{(\lambda(k)+\xi-k)D_+(\xi,z)g(\xi)}{\lambda(\xi)(\xi-k)}\d\xi\,, \\
R_{-,11}=-R_{-,22}=\dashint \frac{D_-(\xi,z)g(\xi)}{\xi-k}\d\xi
\label{e:R-diagonal}
\\
R_{+,12}= R_{+,21}^*=\left\{
\begin{array}{ll}
i\sigma \pi g(k(\zeta))P_+(\zeta,z) & \zeta\in \Real \\
0 & \zeta\in \mathcal{C}\setminus\left\{\pm A\right\}
\end{array}
\right. \\
R_{-,12}= R_{-,21}^*=-i\pi g(k(\zeta))P_-(\zeta,z) \qquad \zeta \in \Real\,.
\end{gather}
\ese
where $R_{\pm,ij}$ denotes the $(i,j)$-th entry of $R_\pm$.
Note that, in~\eref{e:Rpmo,d}, both $\rho_{+,d}$ and $\lambda$ inside the Hilbert transform
take opposite signs on sheets I and II, so the Hilbert transform yields the same result on the two sheets, as it should be.
We conclude that the matrix $R_{+,d}$ is determined  independently of the choice of the integration variable.
Importantly notice that, since $R_\pm(\zeta,z)$ are defined as principal value integrals,
even when they admit extension to the complex $k$-plane, their values are going to be discontinuous across the real $k$-axis.

In general, it is not possible to extend all the entries of $R_\pm$ off the continuous spectrum. Similarly to what happens in the case of symmetric nonzero background in \cite{BGKL}, the off-diagonal parts of $R_\pm$ can be written as
\begin{gather*}
R_{-,o}(\zeta,z)= \pi \lim_{t\to -\infty} \mathcal{H}_k\left[g(\xi)\e^{-i(k-\xi)t\sigma_3}\rho_{-,o}(t,z,\xi)\e^{i(k-\xi)t\sigma_3} \right]\,, \\
R_{+,o}(\zeta,z)= \pi  \lim_{t\to +\infty} \mathcal{H}_k\left[g(\xi)\e^{-i(\lambda(k)-\lambda(\xi))t\sigma_3}\rho_{+,o}(t,z,\xi)\e^{i(\lambda(k)-\lambda(\xi))t\sigma_3}\right]\,.
\end{gather*}
For each matrix element, the Hilbert transform is analytic and bounded in the complex $\zeta$-plane wherever the exponential inside tends to zero
as $t\to \pm \infty$. Hence, looking at the regions where $\Im k \lessgtr 0$  and $\Im \lambda \lessgtr 0$, it follows that
\bse
\label{e:R_pm_offdiagonal}
\begin{gather}
R_{+,12}(\zeta,z)=0 \quad \forall \zeta \in D^+\,,
\qquad
R_{+,21}(\zeta,z)=0\quad \forall \zeta \in D^-\,,
\label{e:R_+_offdiagonal}
\\
R_{-,21}(\zeta,z)=0 \quad \forall \zeta \in \Complex^+\,,
\qquad
R_{-,12}(\zeta,z)=0 \quad \forall \zeta \in \Complex^-\,.
\label{e:R_-_offdiagonal}
\end{gather}
\ese

%%%%%%%%%%%%%%%%%%%%%%%%%%%%%%%%%%%%%%%%%%%%%%%%%%%%%%%%%%%%%%%%%%%%%%%%%%%%%%%%%%%%%%%%%%%%%%%%%%%%%%%%%%%%%%%%%%%%%
\subsection{Propagation of the reflection coefficients}
\label{ss:propagationreflection}

In light of~\eref{e:C_pm}, \eqref{scatt_mat_z} implies
$$
S(\zeta,z)=C_-(\zeta,z)C_+^{-1}(\zeta,z)\,, \qquad \forall \zeta \in \Real\,,
$$
and one can show that
\begin{equation}
\label{e:Sz}
\frac{\partial S}{\partial z}=\frac{i}{2}(R_-S-SR_+)\,, \qquad \forall \zeta \in \Real\,.
\end{equation}
Recall that we have introduced two sets of reflection coefficients, the reflection coefficients from the left,
i.e., $r_-=b/a$ and $r_+=\bar{b}/\bar{a}$, and the reflection coefficients from the right, i.e., $r_+=-\bar{b}/a$ and
$\bar{r}_+=-b/\bar{a}$. In order to obtain the propagation equations for both sets of reflection coefficients, we introduce
$$
B(\zeta,z)=S_o \,(S_d)^{-1}=\begin{pmatrix} 0 & \bar{r}_- \\ r_- & 0 \end{pmatrix}\,, \qquad
\tilde{B}(\zeta,z)=(S_d)^{-1}S_o=-\begin{pmatrix} 0 & r_+  \\ \bar{r}_+ & 0 \end{pmatrix}\,,
$$
and observe that
\begin{gather}
\frac{\partial B}{\partial z}=(S_o)_z(S_o)^{-1}B-B(S_d)_z(S_d)^{-1}\,, \qquad
\frac{\partial \tilde{B}}{\partial z}=\tilde{B}(S_o)^{-1}(S_o)_z-(S_d)^{-1}(S_d)_z\tilde{B}\,.
\end{gather}
Separating \eref{e:Sz} into its diagonal and off-diagonal parts
and substituting into the propagation equations for $B$ and $\tilde{B}$ yields
\bse
\begin{gather}
-2i\frac{\partial B}{\partial z}=R_{-,o}+[R_{-,d},B]-BR_{-,o}B-S_dR_{+,o}(S_d)^{-1}+BS_oR_{+,o}(S_d)^{-1}\,,
\label{e:B_z}
\\
-2i\frac{\partial \tilde{B}}{\partial z}=-R_{+,o}+[R_{+,d},\tilde{B}]+\tilde{B}R_{+,o}\tilde{B}+(S_d)^{-1}R_{-,o}S_d-(S_d)^{-1}R_{-,o}S_o\tilde{B}\,.
\label{e:Btilde_z}
\end{gather}
\ese
First, we express the RHS of~\eref{e:B_z} in terms of the limiting values as $t \to -\infty$. In order to do so, we look at the last three terms in the RHS.
Recall that $R_{\pm,0}$ is given by~\eref{e:Rpmo}, and $\rho_+$ is expressed in terms of $\rho_-$ via~\eref{e:relrho+rho-}. Also, the first symmetry implies that $S^{-1} = S^\dag/\det S$ with $\det S$ given by~\eref{e:detS}. Decomposing \eref{e:relrho+rho-} into its diagonal and off-diagonal parts yields:
\bse
\be
\rho_{+,o} = \frac{1}{\det S}\left(S^\dag_d\, \rho_{-,o}\,S_d + S^\dag_o\,\rho_{-,o}\,S_o + S^\dag_d \,\rho_{-,d}\,S_o + S^\dag_o \,\rho_{-,d}\,S_d  \right)\,.
\ee
Moreover, since $S^\dag S = S S^\dag = (\det S)I$, we have
\be
\label{e:S_relations}
S^\dag_d \,S_d + S^\dag_o \,S_o = S_d \,S^\dag_d + S_o\, S^\dag_o = (\det S) I\,,\quad
S^\dag_d \,S_o + S^\dag_o\, S_d = S_d \,S^\dag_o + S_o\, S^\dag_d = O_{2 \times 2}\,.
\ee
\ese
Substituting the above expressions into the last three terms in the RHS of~\eref{e:B_z}, after simplifications we obtain:
\be
-2i\frac{\partial B}{\partial z}= (1+ \sigma ) R_{-,o}+[R_{-,d},B] - (1-\sigma)\,B R_{-,o}B + i \nu \pi g(\zeta) [\rho_{-,d},B]\sigma_3\,,
\ee
where $\sigma=1$ for $\zeta\in (-\infty,-A]\cup [A,+\infty)$, and $\sigma=-1$ for $\zeta\in (-A,A)$.
Then the propagation equation for $r_-(\zeta,z)$ can be simply obtained from the $(2,1)$-entry of the above matrix equation, namely:
\be
\label{e:propr-}
\partialderiv{r_-}{z}=
\begin{cases}
      -i K^{\mathrm{out}}\,r_- - \pi g(\zeta) P_-^* & \zeta\in (-\infty,-A]\cup [A,+\infty)\,, \\
     -r_- \big[ \pi g(\zeta)\, P_-\,r_- + iK^{\mathrm{in}}\big] & \zeta\in (-A,A)\,,
   \end{cases}
\ee
where
\be
K^{\mathrm{out/in}}(\zeta,z) = \pi \mathcal{H}_k[D_-(\zeta,z)\,g(\zeta)] \pm i\pi D_-(\zeta,z)\,g(\zeta)\,,
\label{e:Kpmdef}
\ee
with $K^{\mathrm{out}}$ and $K^{\mathrm{in}}$ corresponding to the positive and negative signs, respectively, in the right-hand side. Explicitly, we have
\be
\label{e:dr-dz}
\partialderiv{r_-}{z}=
\begin{cases}
      -i \pi \mathcal{H}_k[D_-\,g(\zeta)]\,r_- +\pi g(\zeta) D_-r_-- \pi g(\zeta) P_-^* & \zeta\in (-\infty,-A]\cup [A,+\infty)\,, \\
     -i \pi \mathcal{H}_k[D_-\,g(\zeta)]\,r_--\pi P_- r_-^2 -\pi g(\zeta) D_-r_-  & \zeta\in (-A,A)\,,
\end{cases}
\ee
and one can verify that the propagation equation for $r_-$ is consistent with the symmetry \eref{e:symmr_-}.
Both of the equations in~\eqref{e:dr-dz} can be solved explicitly.
However, to solve \eref{e:propr-} explicitly, it is enough to find $r_-$  when $\zeta \in (-\infty,-A]\cup [A,+\infty)$,
since one can use the symmetry relation \eref{e:symmr_-} to derive $r_-$ in the segment $(-A, A)$.
Solving the linear, non-homogenous equation for $\zeta \in (-\infty,-A]\cup [A,+\infty)$ yields
\bse
\be
r_-(\zeta, z) = \e^{-i \chi(\zeta, z)}\Big[ r_-(\zeta,0) - \pi g(\zeta) \int_0^z P_-^* (\zeta, y) \e^{i \chi(\zeta, y)} \d y\Big] \qquad
\zeta \in (-\infty,-A]\cup [A,+\infty)\,,
\label{e:r-propagation}
\ee
where
\be
\chi(\zeta, z) = \int_0^z K^{\mathrm{out}}(\zeta, y) \d y\,,
\label{e:chidef}
\ee
\ese
with $ K^{\mathrm{out}}$ given in \eqref{e:Kpmdef}. The second reflection coefficient $\=r_-$ can be computed through the first symmetry relation~\eref{e:1stsymmr_-}.

Next, using a similar approach, we express the RHS of~\eref{e:Btilde_z} in terms of the limiting values as $t \to \infty$. Like before, it is useful to write $\rho_{-,o}$ in terms of $\rho_{+}$ as follows:
\be
\rho_{-,o} = \frac{1}{\det S}\left(S_d\, \rho_{+,o}\,S_d^\dag + S_o\,\rho_{+,o}\,S_o^\dag + S_d \,\rho_{+,d}\,S_o^\dag + S_o \,\rho_{+,d}\,S_d^\dag  \right)\,.
\ee
Using the above equation along with~\eref{e:S_relations}, one can rewrite the equation~\eref{e:Btilde_z} as
\be
-2i\frac{\partial \~B}{\partial z}= - R_{+,o}+[R_{+,d},\~B] + \~B R_{+,o}\~B + i\pi g(\zeta) \rho_{+,o}\sigma_3
-i\pi g(\zeta) \~B \rho_{+,o}\~B \sigma_3
+ i\pi g(k) [\~B,\rho_{+,d}]\sigma_3\,.
\ee
Considering the $(1,2)$-component of the above equation, we can find the propagation equation for $r_+(\zeta,z)$.
Specifically, we have
\be
\displaystyle
\partialderiv{r_+}{z}=
\begin{cases}
      -i\Lambda\,r_+ - \pi g(\zeta) P_+ & \zeta\in (-\infty,-A]\cup [A,+\infty)\,,
      \\
     r_+ \big[ \pi g(\zeta)\, P_+^*\,r_+ - i\Lambda\big] & \zeta\in (-A,A)\,,
     \\
     \frac{1}{2}\Big(\pi g(\zeta) P_+^* r_+^2 - 2i \Lambda r_+ - \pi g(\zeta) P_+\Big)
      & \zeta\in \mathcal{C}^+\,,
   \end{cases}
   \label{e:propr_+}
\ee
where
\be
\Lambda = -\pi\lambda(\zeta) \mathcal{H}_k[D_+(\zeta,z)\,g(\zeta)/\lambda(\zeta)] - 2\,w_+(z) -i\pi g(\zeta)D_+(\zeta,z)\,,
\ee
with $w_+(z)$ as in~\eref{e:w_+}.
The above equation can be written as:
\begin{multline}
\partialderiv{r_+}{z}=
i\pi \lambda(\zeta) \mathcal{H}_k[D_+\,g(\zeta)/\lambda(\zeta)]+2iw_+ r_+ -\pi g(\zeta) D_+r_+ \\
+\begin{cases}
     -\pi g(\zeta)P_+ & \zeta\in (-\infty,-A]\cup [A,+\infty)\,,
      \\
    +\pi g(\zeta)P_+^*r_+^2 & \zeta\in (-A,A)\,,
     \\
     +\frac{1}{2}\pi g(\zeta) P_+^* r_+^2 -\pi g(\zeta)P_+  & \zeta\in \mathcal{C}^+\,.
   \end{cases}
\end{multline}
As before, the propagation equation for $\=r_+$ can be obtained using the symmetry relation~\eref{e:1stsymmr_+}.

Next, we verify that the propagation equation for $r_+$ is consistent with the symmetries. First, we differentiate the symmetry relation~~\eref{e:symforr_+} and obtain
\be
\label{e:derivativer_+}
r_+(-A^2/\zeta , z) \partialderiv{r_+(\zeta,z)}{z} + r_+(\zeta , z) \partialderiv{r_+(-A^2/\zeta,z)}{z} = 4i w_+(z)\frac{q_+(0)}{q_+^*(0)}\e^{4iW_+(z)}\,.
\ee
Then we show that ~\eref{e:propr_+} is consistent with the above symmetry. Without loss of generality, suppose $\zeta \in (-A,A)$. Then the LHS of the above equation becomes
\be
\label{e:symderivativer_+}
     r_+(-A^2/\zeta) r_+(\zeta)\big[\pi g(\zeta)\, P_+^*(\zeta)\,r_+(\zeta) - i\Lambda(\zeta) \big]+
    r_+(\zeta)\big[-i \Lambda(-A^2/\zeta) r_+(-A^2/\zeta) - \pi g(-A^2/\zeta) P_+(-A^2/\zeta)\big]\,.
\ee
Using the definition of $\Lambda(\zeta)$ in~\eref{e:propr_+} with equation~\eref{e:D_pm&P_pmsheet1&2} we have the following relations:
\be
\label{e:lambda&P2ndsymm}
\Lambda(-A^2/\zeta,z) = -\Lambda(\zeta,z) - 4w_+(z)\,,
\quad
P_+(-A^2/\zeta,z) =\frac{q_+(0)}{q_+^*(0)}\e^{4iW_+(z)}P_+^*(\zeta,z)\,,
\ee
where $w_+(z)$ and $W_+(z)$ defined in~\eref{e:w_+} and \eref{e:Q_+(z)}, respectively. Combining equations~\eref{e:lambda&P2ndsymm} and \eref{e:symderivativer_+} one can verify the symmetry relation~\eref{e:derivativer_+}.

%%%%%%%%%%%%%%%%%%%%%%%%%%%%%%%%%%%%%%%%%%%%%%%%%%%%%%%%%%%%%%%%%%%%%%%%%%%%%%%%%%%%%%%%%%%%%%%%%%%%%%%%%%%%%%%%%%%%%
\subsection{Propagation of the norming constants}
\label{ss:normingpropagation}

We now derive the propagation equation for the norming constants.
Using the definition of the norming constant $C_n$ with $n=1, \dots, N$ we have
\be
C_n(z) = b_n(z) \lim_{\zeta\to \zeta_n}\dfrac{\zeta - \zeta_n}{a(\zeta,z)}\,.
\label{e:modifiedC_n}
\ee
Differentiating the above equation with respect to $z$ and assuming the limit and the derivative commute, we obtain:
\be
\partialderiv{C_n}{z} = \partialderiv{b_n}{z}\frac{1}{a'(\zeta_n,z)} - C_n(z) \lim_{\zeta\to\zeta_n} \Big(\frac{1}{a(\zeta,z)} \partialderiv{a(\zeta,z)}{z}\Big)\,.
\label{e:C_z}
\ee

As we show next, this is another instance in which the formalism deviates significantly from the case of zero background.
This is because, in order to derive the correct propagation equation for $C_n$ one needs to evaluate \eqref{e:C_z}
by using the expression for $a(\zeta,z)$ obtained using the trace formulae, which are derived in section~\ref{s:trace}.
Specifically, we begin by differentiating~\eref{e:Trace} with respect to $z$, which yields
\be
\partialderiv{a}{z} = \frac{a}{2\pi i} \int_{\Real} \partialderiv{K_1(z,s)}{z}\frac{\d s}{s-\zeta}\, ,\qquad \zeta \in D^+_{\mathrm{out}}\,.
\ee
Using Eqs.~\eref{e:K} and \eref{e:propr-}, one can show the following:
\begin{multline}
\displaystyle
  - \frac{1}{2 \pi}\,\int_{\Real} \partialderiv{K_1(z,s)}{z}\frac{\d s}{s-\zeta}\d s = \int_{(-\infty , -A)\cup (A, \infty)} g(s) \Bigg( \dfrac{D_-(z, s) |\gamma_-(z,s)|^2 - \Re[\gamma_-(z,s)P_-(z,s)]}{1+ |r_-(z,s)|^2} \Bigg) \frac{\d s}{s-\zeta}
   \\
   +
   \int_{(-A , A)} g(s) \Bigg( \dfrac{D_-(z, s)  + \Re[\gamma_-(z,s)P_-(z,s)]}{1+ |r_-(z,s)|^2} \Bigg) \frac{\d s}{s+\zeta}\,.
\end{multline}
One can eliminate $\Re[\gamma_-(z,s)P_-(z,s)]$ using \eref{e:D+P+} and obtain
\be
\label{e:a_ztrace}
\partialderiv{a}{z} = \frac{i}{2}\,\eta(\zeta, z)\,a (\zeta,z)\, ,\qquad \zeta \in D^+_{\mathrm{out}}\,,
\ee
where
\be
\eta(\zeta, z) =
\int_{(-\infty , -A)\cup (A, \infty)} \dfrac{D_-(z, s) - D_+(z, s)}{s - \zeta} g(s)\,\d s
  +\int_{(-A ,A)} \dfrac{D_-(z, s) + D_+(z, s)}{s+ \zeta} g(s)\,\d s\,.
\ee
Performing a change of variable $s \to -A^2/s$ and using the symmetries of $D_\pm$ (Eqs.~\eref{e:D_pm&P_pmsheet1&2}), one can combine the above two integrals and obtain:
\bse
\label{e:eta}
\begin{align}
\eta(\zeta, z)
&=
\int_{(-\infty , -A)\cup (A, \infty)} \left [\dfrac{D_-(z, s) - D_+(z, s)}{s - \zeta} g(s)
  +\dfrac{A^2\,g(-A^2/s) \big(D_-(z, s) - D_+(z, s)\big)}{s(s\,\zeta - A^2)} \right]\,\d s
\notag \\
&=
\int_{(-\infty , -A)\cup (A, \infty)}
\Bigg(\dfrac{1}{s-\zeta} + \dfrac{A^2}{s(s\,\zeta - A^2)} \Bigg)g(s) \big(D_-(z, s) - D_+(z, s)\big)\,\d s
\notag \\
&=
\int_{(-\infty , -A)\cup (A, \infty)}
\dfrac{\zeta (s^2 - A^2)}{s\,(s-\zeta)(s \,\zeta - A^2)} g(s) \big(D_-(z, s) - D_+(z, s)\big)\,\d s\,.
\end{align}
\ese
Now recall from equation~\eref{e:Rpm}, for $z \in \Real$,
\be
\partialderiv{\Phi}{z} = -\frac{i}{2}\Phi R_- + T \Phi\,,
\qquad
\partialderiv{\Psi}{z} = -\frac{i}{2}\Psi R_+ + T \Psi\,.
\ee
Now observe that some columns of the above equations can be extend into the UHP. Namely,
\be
\label{e:phi&psi_z}
\partialderiv{\phi}{z} = -\frac{i}{2}\phi R_{-,22} + T\phi\,,
\qquad
\partialderiv{\psi}{z} = -\frac{i}{2}\psi R_{+,11} + T\psi\,,\quad \Im z>0\,.
\ee
Since at $\zeta = \zeta_n$ one has $\psi(t,z,\zeta_n) = b_n(z)\,\phi(t,z,\zeta_n)$, differentiating with respect to $z$ we obtain
\be
\partialderiv{\psi(\zeta_n)}{z} = \partialderiv{b_n(z)}{z}\phi(\zeta_n) + b_n(z)\partialderiv{\phi(\zeta_n)}{z}\,.
\ee
In turn, using~\eref{e:phi&psi_z}, this yields
\be
\partialderiv{b_n(z)}{z} = -\frac{i}{2}(R_{+,11}(z,\zeta) - R_{-,22}(z,\zeta))\big|_{\zeta=\zeta_n}\,b_n(z)\,.
\label{e:b_z}
\ee
Finally, using Eqs.~\eref{e:b_z},~\eref{e:C_z} and \eref{e:a_ztrace} one can derive the following propagation equation for the norming constants:
\be
\partialderiv{C_n}{z} = -\frac{i}{2} \big(R_{+,11}(\zeta_n) - R_{-,22}(\zeta_n) + \eta(\zeta_n)\big)\, C_n\,,\qquad n=1,\dots, N\,.
\ee
Importantly, \eqref{e:a_ztrace} also shows that the zeros of $a(\zeta)$, i.e., the discrete eigenvalues of the scattering problem,
are independent of~$z$.

It is important to stress that the above derivation of the propagation equations for $a(\zeta,z)$ and for the norming constants does not assume
any analytic continuation beyond what has been established in the direct problem. We show below that the propagation equation that one obtains for $a(\zeta,z)$
assuming that Eq.~\eref{e:R_pm_offdiagonal} can be extended off the real $\zeta$-axis does not appear to coincide with \eref{e:a_ztrace}.

Recall that, for $\zeta \in \Real$, the propagation of the scattering matrix is given by~\eref{e:Sz}. The $(1,1)$ and $(2,1)$ entries of ~\eref{e:R_pm_offdiagonal} yield, respectively,
\be
\partialderiv{a}{z} = \frac{i}{2}\Big( (R_{-,11}-R_{+,11})\,a + b\, R_{-,12} - \=b\, R_{+,21}\Big)\, ,\quad
\partialderiv{b}{z} = \frac{i}{2}\Big((R_{-,22}-R_{+,11})\,b + a\, R_{-,21} - \=a\, R_{+,21}\Big) \,,\quad \zeta \in \Real\,.
\ee
If one assumes that \eref{e:R_pm_offdiagonal} can be extended into an arbitrarily small strip around the real $\zeta$ axis,
one also has:
\be
\partialderiv{a}{z} = \frac{i}{2}( R_{-,11}-R_{+,11})\,a\, ,\quad \zeta \in D^+_{\mathrm{out}}\,.
\qquad
\label{e:a_z}
\ee
In order to compare it with \eref{e:a_ztrace}, note that
\be
\label{e:DiffR_11}
R_{-,11}-R_{+,11} = \dashint\frac{D_-(z, \xi)}{\xi-k} g(\xi) \d\xi - \dashint\frac{\lambda(k)+\xi -k}{\lambda(\xi)\,(\xi-k)} D_+(z, \xi)\,g(\xi) \d\xi\,,
\ee
and the latter has to coincide with $\eta(z)$.
First we perform a variable change $\xi = \frac{1}{2} (s - A^2/s)$, which implies
\be
\dfrac{\d \xi}{\d s} = \frac{1}{2}\Big(1 + \frac{A^2}{s^2}\Big)\,, \qquad
\lambda(\xi) = s -\xi = \frac{1}{2}\Big(s + \frac{A^2}{s}\Big)\,, \qquad
\xi - k = \frac{1}{2}(s-\zeta) \dfrac{(s\,\zeta + A^2)}{s\,\zeta}\,.
\ee
\\
Substituting the above expressions into~\eref{e:DiffR_11} yields
\begin{align}
R_{-,11}-R_{+,11} &=
\int_{(-\infty , -A)\cup (A, \infty)} \Bigg(\dfrac{\zeta (s^2 + A^2)}{s\,(s - \zeta) (s\,\zeta + A^2)} D_-(z,s)\,g(s) -
\dfrac{(s^2 + A^2)\zeta + 2A^2 (s -\zeta)}{s\,(s - \zeta) (s\,\zeta + A^2)} D_+(z,s)\,g(s) \Bigg) \d s
\notag \\
&=
\int_{(-\infty , -A)\cup (A, \infty)}
\dfrac{\zeta (s^2 + A^2)}{s\,(s-\zeta)(s \,\zeta + A^2)} g(s) \big(D_-(z, s) - D_+(z, s)\big) \d s
\notag
\\
&
\kern20em
- 2A^2  \int_{(-\infty , -A)\cup (A, \infty)} \dfrac{g(s) D_+(z,s)}{s\,(s \,\zeta + A^2)} \d s\,.
\end{align}
Comparing the above equation with \eref{e:eta} shows that the two expressions coincide when $A=0$, but not, in general,
in the case of nonzero background. Therefore, when dealing with a nontrivial background, one has to take~\eref{e:a_ztrace} as the correct equation for the propagation of $a(\zeta,z)$.

%%%%%%%%%%%%%%%%%%%%%%%%%%%%%%%%%%%%%%%%%%%%%%%%%%%%%%%%%%%%%%%%%%%%%%%%%%%%%%%%%%%%%%%%%%%%%%%%%%%%%%%%%%%%%%%%%%%%%
\section{Asymptotic states of propagation}
\label{s:lta}

We now show how the IST formalism developed in the previous sections allows one to immediately obtain certain features
about the asymptotic state of the medium as well as information on the asymptotic behavior of the optical pulse in the medium.

\paragraph{Asymptotic value of the scattering coefficients.}
We begin by looking at the asymptotic value of the reflection coefficient for large $z$.
Recall that the evolution (i.e., propagation inside the medium) of the reflection coefficient $r_-(\zeta,z)$ as a function of $z$ is given by~\eqref{e:r-propagation}, with $\chi(\zeta,z)$ given by \eqref{e:chidef} and $K_{\mathrm{out}}(\zeta,z)$ in turn by
\eqref{e:Kpmdef}.
Therefore, its behavior as a function of $z$ is determined by the sign of the imaginary part of $K^{\mathrm{out}}$,
which is given by~\eqref{e:Kpmdef}.
Since $D_-(\zeta,z)$ and $g(\zeta)$ are real-valued, so is the Hilbert transform in \eqref{e:Kpmdef}.
Therefore, $g(\zeta)$ to be non-negative, the growth or decay of $r_-(\zeta,z)$ is completely determined by the sign of $D_-(\zeta,z)$.

Let us consider first the case $P_-(\zeta,z)\equiv0$, since it is the simplest one.
Inspection of~\eqref{e:r-propagation}
shows that if the medium is initially in the stable pure state
(i.e., it is prepared so that $P_-=0$ and $D_-=-1$),
$r_-(\zeta,z)$ is exponentially decaying as $z\to +\infty$ for $\zeta\in (-\infty,-A]\cup [A,+\infty)$,
and exponentially growing for $\zeta \in (-A,A)$.
Conversely, if the medium is initially in the unstable pure state (i.e., it is prepared with $P_-=0$ and $D_-=1$),
then $r_-(\zeta,z)$ is exponentially
growing as $z\to +\infty$ for $\zeta\in (-\infty,-A]\cup [A,+\infty)$, and exponentially decaying for $\zeta \in (-A,A)$.

Finally, it is straightforward to see from \eqref{e:r-propagation} that similar considerations apply when $P_-(\zeta,z)\not\equiv0$.
More precisely, whenever $D_-(\zeta,z)>0$, the reflection coefficient
$r_-(\zeta,z)$ has a similar kind of exponential growth in $z$ as when $D_-(\zeta,z)=1$,
and whenever $D_-(\zeta,z)<0$, $r_-(\zeta,z)$ exhibits exponentially decay to a non-zero value if $P_-\ne0$ .

\paragraph{Asymptotic state of the medium.}

Next we look at the asymptotic state of the medium as $t\to\infty$, as given by $D_+$ and $P_+$,
which are determined by the reflection coefficient $r_-(\zeta,z)$
via~\eref{e:D+P+}.

As discussed at the end of section \ref{s:asympdensity}, 
while the values of $D_-$ and $P_-$ are independent of the sheet chosen,
i.e., whether $\zeta\in(-\infty,-A]\cup[A,\infty)$ or $\zeta\in[-A,A]$, the values of $D_+$ and $P_+$ are different for
$\zeta\in(-\infty,-A]\cup[A,\infty)$ or $\zeta\in[-A,A]$.
At the same time, whenever $P_+\equiv0$ or $P_+\to0$ as $z\to\infty$
[which happens, for example,
when the system is in a pure state, i.e., $P_-\equiv 0$],
one can see from \eqref{e:DPasymptinfty} that the sign of
$D_+$ coincides, at least for sufficiently large $z$, with the sign of $D$ on $\zeta\in(-\infty,-A]\cup[A,\infty)$ or $\zeta\in[-A,A]$.
Therefore, for large $z$ one should consider the value of $D_+$ for $\zeta(-\infty,-A]\cup[A,\infty)$
(i.e., the continuous spectrum on the first sheet) as the physical value.
Hence, to discuss the asymptotic state of a medium initially in a pure state,
it is sufficient to limit ourselves to considering $\zeta\in(-\infty,-A]\cup[A,\infty)$.

Consider first the case in which the medium is initially in the stable pure state
(i.e., $P_-=0$ and $D_-=-1$).
In this case, since $r_-(\zeta,z)$ decays exponentially as $z\to \infty$ for all $\zeta\in(-\infty,-A]\cup[A,\infty)$,
\eqref{e:D+P+} imply that $D_+\to -1$ and $P_+\to 0$ for large $z$.
Therefore, the medium returns to the stable state for sufficiently large propagation distances,
justifying the use of the term ``stable state''.

Conversely, if the medium is initially prepared in the unstable pure state
(i.e., $P_-=0$ and $D_-=1$),
$r_-(\zeta,z)$ is exponentially growing as $z\to \infty$ for all $\zeta\in(-\infty,-A]\cup[A,\infty)$,
and~\eqref{e:DP+} still give $D_+\to -1$ and $P_+\to 0$ for large $z$.
Therefore, the medium reverts to the stable state for sufficiently large propagation distances.
This behavior, which is similar to what happens in the MBE system with ZBG \cite{LiMiller},
may be regarded as a decay process induced by the incident optical pulse.

Finally, the behavior of the reflection coefficient discussed above
also allows us to draw some conclusions when $P_-(z,\zeta)\not\equiv0$.
Namely,
if $D_-(\zeta,z)<0$ one has $D_+(\zeta,z)\to D_-(\zeta,z)$
for $\zeta(-\infty,-A]\cup[A,\infty)$.
Conversely, if $D_-(\zeta,z)>0$ one has $D_+(\zeta,z)\to -D_-(\zeta,z)$
for $\zeta(-\infty,-A]\cup[A,\infty)$.
In both cases, one also has $|P_+(\zeta,z)|\to |P_-(\zeta,z)|$.
There is an important difference with the previous discussion, however:
if the medium is not initially in a pure state [i.e., $P_-(z,\zeta)\not\equiv0$],
\eqref{e:DPasymptinfty}
imply that the behavior of $D(t,z,\zeta)$ as $t\to\infty$ is determined not only by the value of $D_+(z,\zeta)$,
but also by $P_+(z,\zeta)$.

\paragraph{Asymptotic values of the optical pulse.}
We now use the results of the preceding paragraphs to discuss the behavior of the optical pulse inside the medium.

In \cite{LiMiller} it was shown that, in the in the sharp limit, a boundary layer around $z=0$ arises upon propagation.
Specifically, \cite{LiMiller} showed that, for causal solutions, a transition arises over an infinitesimally small propagation
distance (see also the earlier results of \cite{GMZ1983,gabitov84,gabitov85,MN1986,Zakharov80}).
The analysis of the asymptotic behavior of the optical field and medium density matrix in \cite{LiMiller}
also revealed a slow decay of the optical field as $t\to\infty$.
Both results, however, have been established in the sharp-line limit,
and one expects them not to hold when inhomogeneous broadening effects are taken into account.

On the other hand, as we discuss below, inspection of the RHP derived in section~\ref{s:inverse} shows
that two different asymptotic behaviors arise depending
on whether one is considering $z$ near zero or, conversely, the asymptotics at large times with $z$ finite.
To appreciate this dichotomy, recall that the jump matrix~\eqref{e:J1def} that defines
the jump condition~\eqref{e:Jumpcondition1} in the RHP
is expressed in terms of of the reflection coefficient~$r_-(\zeta,z)$ via~\eqref{e:Jodef}.
Therefore, the asymptotic behavior of $r_-(\zeta,z)$ discussed in the above paragraphs
determines the asymptotic behavior of the solutions of the RHP and in turn of those of the MBE.
Specifically,
when $D_-(\zeta,z)<0$,
the behavior of the reflection coefficient
guarantees that the contribution of the radiation to the solution is exponentially decaying as $z\to \infty$.
Therefore, one can expect that, for any finite value of $t$, $q(t,z)\to0$ as $z\to\infty$.

A markedly different scenario arises when $z=0$.
In this case the contribution of the reflection coefficient cannot be ignored,
and the analysis of the RHP must allow one to recover the IC $q(t,0)$ of the problem,
and in particular the boundary conditions $q(t,0)\to0$ as $t\to-\infty$ and
$q(t,0) \to A$ as $t\to\infty$.

Even though a detailed calculation of the long-distance asymptotics of the solutions of the RHP~\eqref{e:Jumpcondition}
is outside the scope of this work, it should be obvious that a transition region must arise to connect the different limits
for $q(t,z)$.

Finally, it is worth mentioning that while the IST has been formulated in terms of the uniformization variable $\zeta$,
the physical variable that measures the deviation of the transition frequency of the atoms from its mean value is $k\in \Real$. All results involving real values of $\zeta$, including the asymptotic states of propagation discussed in this section,
can be rewritten in terms of the physical variable $k$ by replacing $\zeta=k+\sqrt{k^2+A^2}$, the sign of the square root corresponding to choosing the first branch of $\lambda$, i.e., $\zeta\in (\infty,-A)\cup(A,+\infty)$.

%%%%%%%%%%%%%%%%%%%%%%%%%%%%%%%%%%%%%%%%%%%%%%%%%%%%%%%%%%%%%%%%%%%%%%%%%%%%%%%%%%%%%%%%%%%%%%%%%%%%%%%%%%%%%%%%%%%%%
\section{Concluding remarks}
\label{s:conclusions}

In summary, we presented the formulation of the IST for two-level systems with inhomogeneous broadening and one-sided nonzero background.
The formalism combines some features of the IST with zero background to others of the IST with two-sided nonzero background.
We have shown that the reflection coefficient is always nonzero, and therefore no reflectionless solutions exist.
This is similar to what happens in the focusing and defocusing NLS equation with asymmetric NZBG
\cite{PHYSD2016,JMP2014,CM2015}
as well as in the Manakov system with non-parallel NZBG \cite{EAJAM2022}.
As far as the inverse problem is concerned, the specific choice of $2\times2$ matrix for the RHP allows one to
bypass the nonlocality of the jump condition as well as to eliminate the jump across the circle $\mathcal{C}$,
both of which are novel features compared to \cite{CM2015}.

We also briefly discussed the asymptotic behavior of the reflection coefficient for large $z$, and the asymptotic states of the medium and
the limiting values of the optical pulse.
In particular, we showed that if $D_-(z,\zeta)<0$ for all $z$ and for all $k\in\Real$,
the reflection coefficient decays exponentially as $z\to\infty$.
Therefore, for sufficiently large $z$,
the solution becomes effectively reflectionless.
We also showed that, for the kinds of BC considered in this work,
if the initial preparation of the medium is a pure state,
the medium asymptotically tends to the stable pure state.
Finally, we showed that two different asymptotic regimes arise for the optical pulse depending on whether one is considering
the limit $z\to\infty$ with $t$ finite or $t\to\infty$ with $z$ finite.

Note that, in the limit $A\to0$, the formalism of the present work reduces to the one in the case of zero background
in a straightforward way.
Specifically, when $A=0$ one simply has $\lambda(k) = k$ and $\zeta = 2k$;
the branch cut $[-iA,iA]$ shrinks to a single point (the origin $k=0$) and the complex $\zeta$ plane
reduces to the complex $k$ plane up to a factor~2.
The second symmetry (which relates values inside and outside the circle $\mathcal C$) becomes immaterial,
and all integrals in $\zeta$ can be trivially converted to integrals in $k$ (in the principal value sense)
and viceversa.

We reiterate the importance of studying the system in the presence of inhomogeneous broadening,
since this allows us to consider media in arbitrary initial preparations
(i.e., not just pure states), unlike what happens in the sharp line limit,
where the only initial states of the medium that are compatible with the system are the pure ones~\cite{LiMiller}.

The results of this work open up a number of interesting problems for future study.
One such problem is the question of existence and uniqueness of solutions of the RHP.
This question is nontrivial even for the NLS equation.
Indeed, it is known that the RHP for the focusing NLS equation with NZBG does not admit a unique solution
even in the symmetric case, and even when the uniformization variable is used.
In the case of the NLS equation, it was shown in \cite{BilmanMiller} that
one can obtain uniqueness results by formulating the IST
without a uniformization variable and by augmenting the RHP with suitable growth conditions at the branch points.
One can conjecture that the same conditions will also guarantee the uniqueness of solutions for the RHP for the MBE
in the formulation of the IST with the uniformization variable as presented here.
A rigorous analysis of this question remains as a problem for future work.

A related issue is that of the well-posedness of the Cauchy problem for the MBE.
If causality requirement is not imposed, the Cauchy problem for the MBE without inhomogeneous broadening in the initially unstable case was shown to admit multiple non-causal solutions for the same data (and these solutions decay to both stable and unstable pure states as $t\rightarrow +\infty$, see Corollary 4 in \cite{LiMiller}).
Conversely, given a causal incident pulse, there exists at most one causal solution to the MBE problem without inhomogeneous broadening (see Theorem 1 in \cite{LiMiller}).
Causality is also imposed in \cite{Zakharov80} to guarantee uniqueness of solutions of the Gelfand-Levitan-Marchenko equations of the inverse problem (equivalently, this is related to non-uniqueness of solution of the Riemann-Hilbert problem for the eigenfunctions), but the MBE considered in both \cite{LiMiller} and \cite{Zakharov80} are restricted to the sharp-line case.

On the other hand, at present there is no statement about non-well-posedness of the Cauchy problem for the MBE with inhomogeneous broadening (or non-uniqueness of solutions of the GLM equations) in \cite{gabitov85}. It should be noted that the proof of uniqueness of a causal solution provided in \cite{LiMiller} does not rely on integrability, and will likely carry through if inhomogeneous broadening is considered. But this does not necessarily mean that if causality is not imposed, the Cauchy problem for the MBE with inhomoegeneous broadening is ill-posed.
In \cite{Zakharov80}, Zakharov seems to suggest that the non-uniqueness induced by the spontaneous solutions is due to
an arbitrary function analytic in a small neighborhood of the origin,
and that the causality requirement forces this arbitrary function to coincide with the analytic extension of the reflection coefficient.
Again, this seems to be related to the essential singularity at the origin introduced by the sharp-line limit, and might not happen with inhomogeneous broadening.
Moreover, all the above results were established in the case of zero background,
and the presence of NZBG introduces yet another layer of complication.
Based on the above considerations, a study of the well-posedness of MBE with inhomogeneous broadening
in the case of rapidly decaying optical pulses, as well as for pulses on a NZBG is an interesting open problem.

Finally, another obvious and interesting question
concerns a detailed and rigorous study of the ``long time''
--- or more appropriately, in this case, long distance --- behavior of solutions.
This question, and the others above, are left for future work, and
we hope that the results of work will motivate further study on these topics.

\subsubsection*{Acknowledgment}

We thank Sitai Li, Peter Miller and Ildar Gabitov for interesting discussions on these topics.
We also thank the Isaac Newton Institute for Mathematical Sciences for its support and hospitality during the program Dispersive Hydrodynamics when parts of the present work were undertaken.
GB was partially supported by the National Science Foundation under grant DMS-2009487.
BP was partially supported by the National Science Foundation under grant DMS-2106488.

%%%%%%%%%%%%%%%%%%%%%%%%%%%%%%%%%%%%%%%%%%%%%%%%%%%%%%%%%%%%%%%%%%%%%%%%%%%%%%%%%%%%%%%%%%%%%%%%%%%%%%%%%%%%%%%%%%%%%

\subsection*{Data Availability}
Data sharing does not apply to this article as no data sets were generated or analyzed during the
current study.

\medskip
\let\em=\it

\makeatletter
\def\@biblabel#1{#1.}
\small
\def\journal#1#2{\textit{#1}\unskip~\textbf{\ignorespaces #2}}
\def\v#1{\textbf{#1}}

\def\reftitle#1{``#1''}
\let\title=\reftitle

\end{document}